\documentclass[10pt,english,manuscript,nofootinbib]{revtex4-1}
\usepackage[T1]{fontenc}
\usepackage[latin9]{inputenc}
\usepackage[a4paper]{geometry}
\usepackage{babel}
\usepackage{float}
\usepackage{units}
\usepackage{amsmath}
\usepackage{amssymb}
\usepackage{graphicx}
\usepackage{esint}
\usepackage[unicode=true]
 {hyperref}
\usepackage{breakurl}

\makeatletter

\providecommand{\tabularnewline}{\\}

\@ifundefined{textcolor}{}
{%
 \definecolor{BLACK}{gray}{0}
 \definecolor{WHITE}{gray}{1}
 \definecolor{RED}{rgb}{1,0,0}
 \definecolor{GREEN}{rgb}{0,1,0}
 \definecolor{BLUE}{rgb}{0,0,1}
 \definecolor{CYAN}{cmyk}{1,0,0,0}
 \definecolor{MAGENTA}{cmyk}{0,1,0,0}
 \definecolor{YELLOW}{cmyk}{0,0,1,0}
 }

\makeatother

\begin{document}

\title{{\Large A Relationship between the Comoving Particle Number and the
Effective Cosmological Constant}}

\thanks{\emph{In memory of my CoCo. }}

\author{Yu-chung Chen}

\email{yuchung.chen@gmail.com}

\selectlanguage{english}%

\address{\mbox{Department of Physics, National Taiwan University;} \\ \mbox{Taipei 10617, Taiwan, R.O.C.}}
\begin{abstract}
In order to discuss and obtain the remaining inflaton potential, we
introduced an idea called ``effective static friction'' in our last
paper \cite{YCCHEN} to balance the ``force'', $\left.\frac{dV}{d\phi}\right|_{\phi=\phi\left(t_{rest}\right)}$,
of inflaton. According to this idea, we now discover that, after the
course of particle creation, there will be a relationship between
the final relativistic particle number\emph{ }inside an arbitrary
chosen comoving volume ($N_{\mathrm{r}}\left(t_{rest}\right)$) and
the effective cosmological constant ($\Lambda$) in our Universe.
This relationship can be expressed as $N_{\mathrm{r}}\left(t_{rest}\right)=\left(\frac{\ell^{2}m_{\phi}^{2}\Lambda}{4\pi G\sigma^{2}\left(t_{rest}\right)}\right)^{\nicefrac{3}{2}}$
when we employ the classical chaotic model, $V\left(\phi\right)=\frac{1}{2}m_{\phi}^{2}\phi^{2}$,
and consider that $\phi$ comes to rest at $t_{rest}$. Moreover,
we obtain an evolution equation for the particle number ($N_{\mathrm{r}}\left(t\right)$)
inside the comoving volume. Meanwhile, a new inflaton field equation
which contains parameters of $N_{\mathrm{r}}\left(t\right)$ and ``particle
creation coefficients'' can also be found. Importantly, the results
illustrate the fact that $\Lambda$ and $N_{\mathrm{r}}$ are the
results of probability.
\end{abstract}

\keywords{accelerating expansion, cosmological constant, comoving entropy,
comoving particle number, inflaton, inflation, vacuum energy density.}

\maketitle
\tableofcontents{}

\noindent \rule[0.5ex]{1\columnwidth}{0.5pt}

\section{Introduction \label{sec:Introduction}}

Since Einstein introduced the cosmological term ($\lambda$) in 1917
\cite{Einstein}, several works have been proposed around the topic.
However, including the research and observations of Hubble et al.
\cite{Friedman,Hubble,Lemaitre}, many of them have rejected the need
for it as a requirement of cosmology. Nevertheless, the term's existence
is still an issue because vacuum energy density has been discovered
in studies on quantum field theory \cite{Zel'dovich}. Unfortunately,
this still fails to provide a solution, as evidenced by the profound
awareness of Weinberg, who indicates that vacuum energy densities
can not be candidates for the cosmological term \cite{Weinberg-1}.

Interestingly, the term, which has a new role as the effective cosmological
constant (ECC, $\Lambda$), became an active player again due to the
amazing observations proposed by Riess and Schmidt et al. in 1998
\cite{acc:1998} and Perlmutter et al. in 1999 \cite{acc:1999}. These
oppose intuition, revealing that our Universe is expanding with acceleration.
Whereafter, since observational data draws out the famous tiny $\Lambda$
problem %
\footnote{Weinberg indicated the following vacuum energy densities in \cite{Weinberg-1}:
Planck, $\left\langle \varepsilon_{\mathrm{Planck}}\right\rangle \approx\unit[10^{115}]{eV^{4}}$;
spontaneous symmetry breaking (SSB) in electroweak (EW) theory, $\left\Vert \left\langle \varepsilon_{\mathrm{EW}}\right\rangle _{\mathrm{SSB}}\right\Vert \approx\unit[10^{44}]{eV^{4}}$;
quantum chromodynamics (QCD), $\left\langle \varepsilon_{\mathrm{QCD}}\right\rangle _{\mathrm{vac}}\approx\unit[10^{32}]{eV^{4}}$.
In addition, he noted the huge difference between vacuum energy densities
and dark energy density, resulting from the discovery that the density
of dark energy is $\varepsilon_{\mathrm{DE}}\simeq\unit[7.17\times10^{-9}]{eV^{4}}$
\cite{Weinberg-2}.%
}, we are immersed in confusion again.

On the other hand, Guth's excellent research \cite{Guth} indicates
a new scenario---inflationary theory---that enables our Universe to
solve the problems which emerge when observations are made using Hot
Big Bang theory %
\footnote{\label{fn:HBB-problems} These problems are the homogeneous, isotropic,
horizon, flatness, initial perturbation, magnetic monopole, total
mass, total entropy and so on. Further reading in \cite{Linde}.%
}. Accordingly, vacuum energy density demands attention once again,
because it needs to be of a huge value in order to initially trigger
inflation. Due to these works, we are now aware of a bigger picture
that illustrates how the Universe (smoothly) ends inflation. Furthermore,
other outstanding research---the reheating \cite{reheating-1,reheating-2,reheating-3}
and warm inflation \cite{Berera et al.} scenarios ---provides detailed
discussion-material for questions relating to the course of the creation
of matter in the very early Universe.

As is evident from the history of physics, simplicity and continuity
are usually required indications of success for new theories. However,
even though the simplest explanations in this case are that the cosmological
constant can play the role of a fundamental constant in the equation
of general relativity, or be a nonzero minimum potential of some scalar
field(s), there remain coincidental problems that cannot be abandoned
\cite{YCCHEN}. Therefore we believe that the simplest explanation
for accelerating expansion is the remaining inflaton potential. Based
on these beliefs and the knowledge of mechanics, we are able to suggest
a guess---the ``effective friction''---to balance the ``force''
dependent on the remaining inflaton potential \cite{YCCHEN}. Truthfully
speaking, this is not an entirely fresh idea because similar thoughts
have been mentioned in the warm inflation theory of Berera, Fang,
Moss and others \cite{Berera et al.,Moss}. In particular, their work
has analyzed the ``dissipation term %
\footnote{\label{fn:field-equation-WI} The field equation of warm inflation
is proposed as $\ddot{\phi}+\left(3H+\Upsilon\right)\dot{\phi}+V^{'}\left(\phi\right)=0.$
$\Upsilon\dot{\phi}$ is the dissipation term.%
}''---a kind of damping force of inflaton---in the interaction of
fields \cite{dissipation 1,dissipation 2}.

If our conjecture about effective friction proves feasible, an ``effective
static frictional force'' is actually needed for a produced and fixed
cosmological constant. Nevertheless, such a force is very difficult
to imagine in field theory and only leads to the discussion of logic
and phenomenon outlined in our last paper \cite{YCCHEN}.

However, a relationship between $N_{\mathrm{r}}\left(t_{rest}\right)$
(the final particle number of radiation \emph{inside an arbitrary
chosen comoving volume}, the final comoving particle number or FCPN
for short) and $\Lambda$ (ECC) has been found, even if it is only
obtained through the \textquotedblleft{}phenomenon\textquotedblright{}
of inflaton dynamics. (Here we use the term ``particle number''
to describe the \emph{sum of the lepton, baryon (quark) and gauge
boson} \emph{number}s. However, these cannot be gauged individually.)
As mentioned in the abstract, this is $N_{\mathrm{r}}\left(t_{rest}\right)=\left(\frac{\ell^{2}m_{\phi}^{2}\Lambda}{4\pi G\sigma^{2}\left(t_{rest}\right)}\right)^{\nicefrac{3}{2}}$.
What is more, we can provide a numerical result that approaches to
the current observational estimates \cite{entropy 1,entropy 2,BND_1997,WMAP7}.

Now, we would like to introduce our consideration in the following
context. The structure of this paper is as follows: Firstly, in Section
\ref{sec:Effective-friction} we will indicate the relationship of
energy transfer between inflaton and radiation during the inflation
course. The discussion of effective friction will be reviewed later.
In Section \ref{sec:CPN-Lambda}, besides proposing a relationship
between $N_{\mathrm{r}}\left(t_{rest}\right)$ and $\Lambda$, we
will also present a discussion about the essence of $N_{\mathrm{r}}\left(t_{rest}\right)$
and the relationship between $N_{\mathrm{r}}\left(t_{rest}\right)$
and the corresponding entropy obtained from a chosen comoving volume.
Meanwhile, numerical results will be shown in Table \ref{tab:N_vs_Lamda}.
In Section \ref{sec:prob_L-PCC}, we will provide a consistent illustration
for the coefficient of radiation creation as defined in Section \ref{sec:CPN-Lambda}.
Employing the discovery of the coefficient's structure, we gain the
following three benefits: 1. a new inflaton field equation; 2. a reasonable
explanation for the expectation that $\Lambda$ and $N_{\mathrm{r}}\left(t_{rest}\right)$
are the probabilistic productions; 3. the discovery of the courses
that the particle number will be decreased at some stages. Based on
these benefits, a further discussion about $N_{\mathrm{r}}\left(t\right)$
will be proposed. Finally, we will offer conclusions and discussions
in Section \ref{sec:Conclusion}. 

In addition, a toy example to depict our findings will be proposed
in Appendix \ref{apx:PC-example}. The pictures given here not only
illustrate the evolution of the comoving particle number, but also
forge an understanding of the relationship between the motion of inflaton
and the reversed effective kinetic friction.

For convenience, the Natural units: $c=k_{\mathrm{B}}=\hbar=1$ are
used through this paper. The definitions and illustrations of needed
symbols are listed in Table \ref{tab:symbol-definition} on the last
page. We strongly suggest the reader to peruse this table prior to
beginning the paper as it will aid with later discussions.

\section{Effective friction of inflaton dynamics \label{sec:Effective-friction}}

\subsection{Energy transfer between inflaton and radiation \label{sub:energy-transfer}}

First of all, we should illustrate the models and conditions for describing
the energy transfer between inflaton ($\phi$, which is set as a real
scalar field) and radiation ($\mathrm{r}$, the relativistic particles)
\emph{during and after} the epoch of inflation. We assume that our
Universe was also homogeneous and isotropic at the very early age.
Therefore, the FRW line element %
\footnote{\label{fn:definition-R} $\mathbb{R}$ is the spatial scale factor.
We define $\mathbb{R}\left(t_{\mathrm{now}}\right)=1$. The adopted
comoving coordinates are $\left[x^{\mu}\right]=\left(t,\, r,\,\theta,\,\varphi\right)$.
$\mathrm{k}$ is the curvature parameter that takes values of $-1$
(pseudo-3-sphere), $0$ (flat space) or $+1$ (3-sphere) for the geometry
of a Universe.%
} 
\begin{equation}
ds^{2}=dt^{2}-\mathbb{R}^{2}\left(t\right)\left(\frac{dr^{2}}{1-\mathrm{k}r^{2}}+r^{2}d\theta^{2}+r^{2}\sin^{2}\theta d\varphi^{2}\right)\label{eq:2.1}
\end{equation}
should be employed to provide the spacetime background for the equation
of general relativity, as 
\begin{equation}
R_{\mu\nu}-\frac{1}{2}Rg_{\mu\nu}=-8\pi G\left(T_{\mu\nu}^{\left(\phi\right)}+T_{\mu\nu}^{\left(\mathrm{r}\right)}\right).\label{eq:2.2}
\end{equation}
Here, we order equation \eqref{eq:2.2} to contain inflaton and radiation
but without the cosmological constant/term. For the sake of simplicity
and consistency with \eqref{eq:2.1}, $T_{\mu\nu}^{\left(\phi\right)}$
(the inflaton energy-momentum tensor) and $T_{\mu\nu}^{\left(\mathrm{r}\right)}$
(the radiation energy-momentum tensor) should be off-diagonal. $\phi\left(x^{\mu}\right)=\phi\left(t\right)$
guarantees this off-diagonal and provides components of $T_{\mu\nu}^{\left(\phi\right)}$,
\begin{equation}
g_{jj}^{-1}\begin{array}{l}
T_{00}^{\left(\phi\right)}=\varepsilon_{\phi}\left(t\right)=\frac{1}{2}\dot{\phi}^{2}+V\left(\phi\right),\\
T_{jj}^{\left(\phi\right)}=p_{\phi}\left(t\right)=\frac{1}{2}\dot{\phi}^{2}-V\left(\phi\right),\\
T_{\alpha\beta}^{\left(\phi\right)}=0,\;\left(\alpha\neq\beta\right),
\end{array}j=1,\,2,\,3\label{eq:2.3}
\end{equation}
when we employ $T_{\mu\nu}^{\left(\phi\right)}=\partial_{\mu}\phi\partial_{\nu}\phi-g_{\mu\nu}\left(\frac{1}{2}g^{\alpha\beta}\partial_{\alpha}\phi\partial_{\beta}\phi-V\left(\phi\right)\right)$.
Here we define the minimum value of inflaton potential ($V\left(\phi\right)$)
as zero. Besides, due to the form of the radiation energy-momentum
tensor,
\begin{equation}
T_{\mu\nu}^{\left(\mathrm{r}\right)}=\frac{\varepsilon_{\mathrm{r}}\left(t\right)}{3}\left(4u_{\mu}u_{\nu}-g_{\mu\nu}\right)\label{eq:2.4}
\end{equation}
(owing to the pressure of radiation $p_{\mathrm{r}}=\nicefrac{\varepsilon_{\mathrm{r}}}{3}$),
the 4-velocity of radiation $u_{\mu}$ must obviously be
\begin{equation}
\left[u_{\mu}\right]=\left(1,\,0,\,0,\,0\right)\label{eq:2.5}
\end{equation}
to obey the requirement that our Universe has no net matter-current
on average.

The following two facts are worthy of note: Firstly, the conditions
of $T_{\mu\nu}^{\left(\phi\right)}$ and $T_{\mu\nu}^{\left(\mathrm{r}\right)}$
are consistent with a perfect fluid which has no net current, i.e.,
the energy-momentum tensor is
\begin{equation}
T_{\mu\nu}=\left(\varepsilon+p\right)u_{\mu}u_{\nu}-pg_{\mu\nu},\label{eq:EMT}
\end{equation}
and the 4-velocity of elements should be $\left[u_{\mu}\right]=\left(1,\,0,\,0,\,0\right)$.
Since $u_{k}=0$ ($k=1,\,2,\,3$), we  regard it as a condition that
the elements must be static at the comoving coordinates. Then, we
can combine the above settings to write down the Friedman equations
with field $\phi$ and radiation as
\begin{equation}
\frac{\ddot{\mathbb{R}}}{\mathbb{R}}=-\frac{8\pi G}{3}\left(\dot{\phi}^{2}-V\left(\phi\right)+\varepsilon_{\mathrm{r}}\right),\label{eq:Friedman-1}
\end{equation}
\begin{equation}
\left(\frac{\dot{\mathbb{R}}}{\mathbb{R}}\right)^{2}=\frac{8\pi G}{3}\left(\frac{1}{2}\dot{\phi}^{2}+V\left(\phi\right)+\varepsilon_{\mathrm{r}}\right)-\frac{\mathrm{k}}{\mathbb{R}^{2}}.\label{eq:Freidman-2}
\end{equation}

Since we believe that \textbf{our Universe is unique or adiabatic},
the materials inside it should satisfy the conservation law of $D_{\mu}\left(T_{\left(\mathrm{r}\right)}^{\mu\nu}+T_{\left(\phi\right)}^{\mu\nu}\right)=0$
(where $D_{\mu}$ is the covariant derivative). For $\nu=0$, the
law becomes the form of energy conservation, as 
\begin{equation}
\dot{\varepsilon}_{\mathrm{r}}+4H\varepsilon_{\mathrm{r}}=-\left[\dot{\varepsilon}_{\phi}+3H\left(\varepsilon_{\phi}+p_{\phi}\right)\right].\label{eq:2.6}
\end{equation}
Clearly, this shows the relationship of energy transfer between $\phi$
and radiation. Based on the considerations of \cite{Thermal_inflation-1,Thermal_inflation-2}
and \cite{Berera et al.}, we now introduce two postulates:
\begin{description}
\item [{Postulate\ A}] \emph{There exists some interaction between $\phi$
and radiation.} In other words, we can bring the interaction term
$Q\left(t\right)$ into \eqref{eq:2.6} to have 
\begin{equation}
\dot{\varepsilon}_{\mathrm{r}}+4H\varepsilon_{\mathrm{r}}=Q\left(t\right),\quad\dot{\varepsilon}_{\phi}+3H\left(\varepsilon_{\phi}+p_{\phi}\right)=-Q\left(t\right).\label{eq:2.7}
\end{equation}
It follows that $\phi$ and radiation are open to each other.
\item [{Postulate\ B}] \emph{Radiation should be created continuously
during and after the epoch of inflation. }The particle creation course
for our adiabatic Universe can be regarded as a self-heating system.
We now temporarily separate our Universe into a Real part and an Imaginary
part. Meanwhile, we put the created particles into the ``Real Universe'',
and the ``heater'' into the ``Imaginary Universe''. It is very
easy to see that the entropy of the Real will increase during the
course of particle creation, i.e., the Gibbs equation for the ``Real
Universe'' is 
\begin{equation}
TdS_{\mathrm{r}}^{Re}=d\left(\varepsilon_{\mathrm{r}}v\right)+p_{\mathrm{r}}dv>0.\label{eq:Gibbs-real}
\end{equation}
However, if we combine the Real with the Imaginary, the Gibbs equation
should be written as
\begin{equation}
Td\left(S_{\mathrm{r}}^{Re}+S_{\mathrm{r}}^{Im}\right)=d\left(\varepsilon_{\mathrm{r}}v\right)+p_{\mathrm{r}}dv-\overline{dQ_{\mathrm{r}}}=0\label{eq:Gibbs-radiation}
\end{equation}
due to the energy, $\overline{dQ_{\mathrm{r}}}$, output from the
``heater''. Assuredly, the Real and the Imaginary are open to each
other. For this reason, Prigogine et al. propose a description/illustration
of the heating process \cite{Prigogin}. They suggest the thermal
condition of an open radiation-system in an expanding adiabatic Universe,
as 
\begin{equation}
d\left(\varepsilon_{\mathrm{r}}v\right)+p_{\mathrm{r}}dv-\frac{h_{\mathrm{r}}}{n_{\mathrm{r}}}d\left(n_{\mathrm{r}}v\right)=0,\label{eq:2.8}
\end{equation}
which obeys the first law of thermodynamics. Here, $h_{\mathrm{r}}=\varepsilon_{\mathrm{r}}+p_{\mathrm{r}}=\frac{4}{3}\varepsilon_{\mathrm{r}}$;
$v=\frac{4}{3}\pi\left(\mathbb{R}\left(t\right)\ell\right)^{3}$ is
the chosen comoving volume ($\ell$ is a fixed comoving coordinate
distance %
\footnote{\label{fn:coordinate-distance} Since light moves along null geodesics
($ds^{2}=0$), the coordinate distance that it has peregrinated from
point $A$ to point $B$ should be
\begin{equation}
\ell_{A\rightarrow B}=\int_{r_{A}}^{r_{B}}\frac{dr}{\sqrt{1-kr^{2}}}=\int_{t_{start}}^{t_{end}}\frac{dt}{\mathbb{R}\left(t\right)}.\label{eq:coord-distance}
\end{equation}
Here we adopt $d\theta=d\phi=0$ for the path. Dependent on \eqref{eq:coord-distance},
the comoving distance at any cosmic time should be $\mathbb{R}\left(t\right)\ell$
for the chosen coordinate distance $\ell$. A further illustration
can be seen in Figure \ref{fig:FPNo}.%
}); $n_{\mathrm{r}}\left(t\right)=\nicefrac{N_{\mathrm{r}}\left(t\right)}{v\left(t\right)}$
is the number density of relativistic particles; and $N_{\mathrm{r}}\left(t\right)$
is the particle number (\emph{sum of the lepton, baryon (quark) and
gauge boson} \emph{number}s) inside a comoving volume (we call it
the comoving particle number, or CPN, for short). Clearly, \eqref{eq:2.8}
has a consistent form as 
\begin{equation}
\dot{\varepsilon}_{\mathrm{r}}+4H\varepsilon_{\mathrm{r}}=\frac{4}{3}\Gamma\varepsilon_{\mathrm{r}},\label{eq:2.9}
\end{equation}
where $\Gamma\left(t\right)\equiv\nicefrac{\dot{N}_{\mathrm{r}}}{N_{\mathrm{r}}}$
is the ratio of particle creation (RPC). Then, the solution of radiation
energy density $\varepsilon_{\mathrm{r}}$ is 
\begin{equation}
\varepsilon_{\mathrm{r}}\left(t\right)=\varepsilon_{\mathrm{r}}\left(t_{i}\right)\exp\left[\frac{4}{3}\int_{t_{i}}^{t}\left(\Gamma-3H\right)d\tau\right].\label{eq:2.10}
\end{equation}
Alternatively, the CPN at time $t$ during the radiation-dominated
era can be shown as 
\begin{equation}
N_{\mathrm{r}}\left(t\right)=\chi\left[\varepsilon_{\mathrm{r}}\left(t\right)\left(\mathbb{R}\left(t\right)\ell\right)^{4}\right]^{\nicefrac{3}{4}}.\label{eq:2.11}
\end{equation}
Due to the reason of $d\left(S_{\mathrm{r}}^{Re}+S_{\mathrm{r}}^{Im}\right)=0$
during all of our Universe\textquoteright{}s evolutionary stage, the
maximum $S_{\mathrm{r}}^{Re}+S_{\mathrm{r}}^{Im}$ provides the information
that the total parts of the energy transfer of radiation is the state
of the thermal equilibrium. Given that the particles of radiation
can be divided into bosons and fermions, the dimensionless integral
constant $\chi$, which is dependent on initial, can also be found
as
\begin{equation}
\chi=\frac{n_{\mathrm{b}}\left(t\right)+n_{\mathrm{f}}\left(t\right)}{\left[\varepsilon_{\mathrm{b}}\left(t\right)+\varepsilon_{\mathrm{f}}\left(t\right)\right]^{\nicefrac{3}{4}}}=\frac{\frac{\zeta\left(3\right)}{\pi^{2}}\left(\underset{\mathrm{b}}{\sum}g_{\mathrm{b}}T_{\mathrm{b}}^{3}\left(t\right)+\frac{3}{4}\underset{\mathrm{f}}{\sum}g_{\mathrm{f}}T_{\mathrm{f}}^{3}\left(t\right)\right)}{\left[\frac{\pi^{2}}{30}\left(\underset{\mathrm{b}}{\sum}g_{\mathrm{b}}T_{\mathrm{b}}^{4}\left(t\right)+\frac{7}{8}\underset{\mathrm{f}}{\sum}g_{\mathrm{f}}T_{\mathrm{f}}^{4}\left(t\right)\right)\right]^{\nicefrac{3}{4}}}\label{eq:2.12}
\end{equation}
according to the Bose-Einstein and Fermi-Dirac distributions. Here
$\zeta\left(3\right)=1.20206\ldots$ is the Riemann zeta function
of $3$; $g_{\mathrm{b}}$ and $g_{\mathrm{f}}$ are degrees of freedom
for bosons and fermions; and $T_{\mathrm{b}}\left(t\right)$ and $T_{\mathrm{f}}\left(t\right)$
are the temperatures of bosons and fermions. In addition, $T_{\mathrm{b,\, f}}\left(t\right)\gg m_{\mathrm{b,\, f}}$
and $T_{\mathrm{b,\, f}}\left(t\right)\gg\mu_{\mathrm{b,\, f}}$ during
the radiation-dominated era (where $m$ is the mass of particle, and
$\mu$ is the chemical potential). Moreover, due to the structure
of equations \eqref{eq:2.8} and \eqref{eq:2.12}, \emph{we can not
distinguish the amount of each species from equation} \eqref{eq:2.11}\emph{
in our scenario}.
\end{description}
In the similar way, the Gibbs equation of $\phi$ for the ``combined''
Universe is
\begin{equation}
Td\left(S_{\phi}^{Re}+S_{\phi}^{Im}\right)=d\left(\varepsilon_{\phi}v\right)+p_{\phi}dv+\overline{dQ_{\phi}}=0.\label{eq:Gibbs-phi}
\end{equation}
We find that $\frac{\overline{dQ_{\phi}}}{dt}v^{-1}\left(t\right)$
is equal to $Q\left(t\right)=\frac{4}{3}\Gamma\varepsilon_{\mathrm{r}}$
because of equation \eqref{eq:2.6}---the energy conservation relation.
Now, going back to equation \eqref{eq:2.9}, since $\Gamma$ should
not be less than zero (according to Postulate B), the interaction
term $Q\left(t\right)=\frac{4}{3}\Gamma\varepsilon_{\mathrm{r}}\geq0$
concludes from \eqref{eq:2.7} that the energy of $\phi$ will decrease
with time and flow into radiation creation. In other words, inflaton
$\phi$ is somewhat the heater for creating radiation.

\subsection{Inflaton's effective friction \label{sub:Inflaton's EFF}}

Return to equation \eqref{eq:2.6}. After we take the components of
\eqref{eq:2.3} and the condition of radiation creation of \eqref{eq:2.9}
into \eqref{eq:2.6}, the field equation of $\phi$ will be obtained
as 
\begin{equation}
\dot{\phi}\ddot{\phi}+3H\dot{\phi}^{2}+V^{'}\left(\phi\right)\dot{\phi}=-\frac{4}{3}\Gamma\varepsilon_{\mathrm{r}}.\label{eq:2.13}
\end{equation}
The term of $-\frac{4}{3}\Gamma\varepsilon_{\mathrm{r}}$ can be regarded
as the power resulting from some ``kinetic frictional force'' in
phenomenon. Dividing $\dot{\phi}$ into both sides of the equal sign
in \eqref{eq:2.13}, the equation of motion of $\phi$ ($\phi$-EOM
for short) is written as 
\begin{equation}
\ddot{\phi}+3H\dot{\phi}+V^{'}\left(\phi\right)=-\frac{4\Gamma\varepsilon_{\mathrm{r}}}{3\dot{\phi}}.\label{eq:2.14}
\end{equation}
Undoubtedly, 
\begin{equation}
\mathbf{f}_{\phi k}\equiv-\frac{4\Gamma\varepsilon_{\mathrm{r}}}{3\dot{\phi}}\label{eq:2.15}
\end{equation}
can be called the ``effective kinetic frictional force'' (EKFF)
for the oscillating system of $\phi$ (the $\phi$-system for short)
because its direction, $\left\rangle \mathbf{f}_{\phi k}\right\langle $,
is indeed opposite to $\left\rangle \dot{\phi}\right\langle $. In
addition, from the fact of \eqref{eq:2.13}, $\Gamma$ will be zero
(ceasing to create radiation) when $\dot{\phi}=0$. Therefore, we
can audaciously guess the amount of ``effective static frictional
force'' (ESFF) as

\begin{equation}
\left\Vert \boldsymbol{\mathbf{f}}_{\phi s}\right\Vert =\underset{\dot{\phi}\rightarrow0}{\lim}\left\Vert -\frac{4\Gamma\varepsilon_{\mathrm{r}}}{3\dot{\phi}}\right\Vert .\label{eq:2.16}
\end{equation}
It should be noted that, there is no problem with equation \eqref{eq:2.16}
because \textbf{no work will be done by a static frictional force}.

\section{The comoving particle number and the effective cosmological constant \label{sec:CPN-Lambda}}

\subsection{The $\Lambda$-CPN relationship \label{sub:Lambda-No.-relationship}}

In this section, we would like to discuss effective friction in greater
depth. Because RPC is defined as $\Gamma=\frac{\dot{N}_{\mathrm{r}}}{N_{\mathrm{r}}}$,
the evolution of radiation energy density can be alternatively rewritten
as
\begin{eqnarray}
\mathbf{f}_{\phi k}\cdot\dot{\phi}\left(t\right) & = & -\frac{4\Gamma\varepsilon_{\mathrm{r}}}{3}\nonumber \\
 & = & -\frac{4}{3}\frac{\dot{N}_{\mathrm{r}}\left(t\right)}{N_{\mathrm{r}}\left(t\right)}\left(\frac{N_{\mathrm{r}}\left(t\right)}{\chi}\right)^{\nicefrac{4}{3}}\left(\mathbb{R}\left(t\right)\ell\right)^{-4}\label{eq:3.1}
\end{eqnarray}
when we take \eqref{eq:2.11} into \eqref{eq:2.15}. According to
the conclusion of \cite{Prigogin}, the change of the particle number
in equation \eqref{eq:2.8} is due to the energy transfer from gravitation
(the expansion or shrinking of a Universe) to matter. We can logically
suppose the particle creation rate as 
\begin{equation}
\dot{N}_{\mathrm{r}}\left(t\right)=\alpha\left(t\right)\ell^{3}\mathbb{R}^{4}\left(t\right)\cdot\dot{\phi}\left(t\right),\label{eq:3.2}
\end{equation}
where $\alpha\left(t\right)$ is the coefficient of radiation creation
(CRC). Then, the EKFF can be expressed as
\begin{equation}
\mathbf{f}_{\phi k}=-\frac{4\alpha\left(t\right)}{3\chi^{\nicefrac{4}{3}}}\ell^{-1}N_{\mathrm{r}}^{\nicefrac{1}{3}}\left(t\right)\equiv-\sigma\left(t\right)\ell^{-1}N_{\mathrm{r}}^{\nicefrac{1}{3}}\left(t\right).\label{eq:EFF}
\end{equation}
Here $\sigma\left(t\right)$ is defined as the reduced coefficient
of radiation creation (RCRC), which is proportional to the CRC. It
is also notable that the units of CRC and RCRC are $\left[\alpha\left(t\right)\right]=\left[\sigma\left(t\right)\right]=\unit{s^{-2}}$.
Moreover, the sign of $\sigma\left(t\right)$ (or $\alpha\left(t\right)$)
is the same as $\dot{\phi}$ (in the following context, we use ``$\left\rangle \sigma\left(t\right)\right\langle =\left\rangle \dot{\phi}\right\langle $''
for short as the description of the situation) to follow the fact
of particle creation. In general, the CRC or RCRC is considered to
be a function of time. Of course, it also includes the possibility
that $\left\Vert \sigma\right\Vert $ is a constant. In actual fact,
\eqref{eq:3.2} asserts another conclusion: \emph{the stopping $\phi$
can be employed to answer the question of why we have never seen a
particle created spontaneously even though the Universe is expanding.}

\begin{figure}
\caption{\label{fig:FPNo}Due to the homogeneous and isotropic properties of
our Universe, the particle number inside a chosen comoving volume
should be a constant after the creation has been ceased. For the two
pictures below, meanings are signified by colorized objects: the blue
line is the comoving circle; red points are the particles inside the
circle; green points are particles outside the circle; grey lines
are the comoving coordinates. The left hand picture shows a younger
Universe at $t_{0}$ old, and the chosen comoving circle has a comoving
radius of $\mathrm{a}=\mathbb{R}\left(t_{0}\right)\ell$. The right
hand picture is the same Universe but its age is $t_{0}+\Delta t$,
and the comoving radius enlarges to $\mathrm{b}=\mathbb{R}\left(t_{0}+\Delta t\right)\ell$.
Here, $\ell$ denotes the chosen comoving coordinate radius.}

\centering{}\includegraphics[scale=0.5]{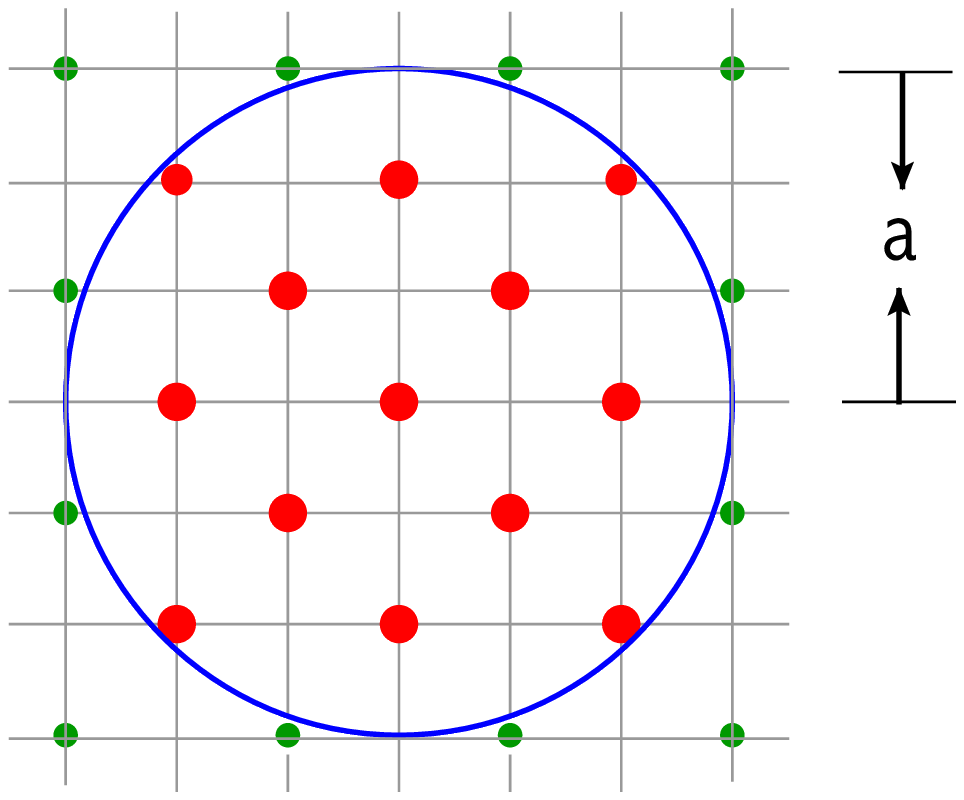}$\quad$\includegraphics[scale=0.9]{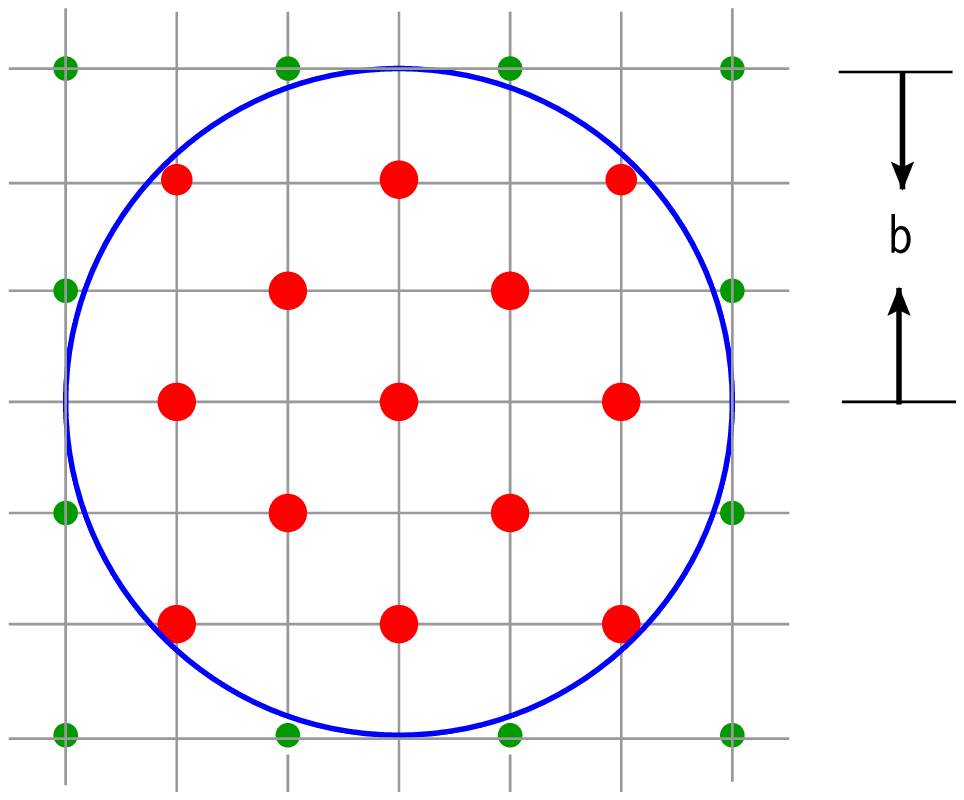}
\end{figure}

Now, suppose that an ESFF exists in the $\phi$-system. This leads
to the expectation that $\phi$ will finally come to rest at the position
of $\phi\left(t_{rest}\right)$. It is actually possible that $\phi\left(t_{rest}\right)$
is not the position of minimum point of $V\left(\phi\right)$ \cite{YCCHEN}.
Correspondingly, the following objects will become fixed when $t\geq t_{rest}$,
as 
\begin{eqnarray}
N_{\mathrm{r}}\left(t\right) & = & N_{\mathrm{r}}\left(t_{rest}\right),\label{eq:FPNo}\\
V\left(\phi\left(t\right)\right) & = & V\left(\phi\left(t_{rest}\right)\right).\label{eq:3.3}
\end{eqnarray}
Here $N_{\mathrm{r}}\left(t_{rest}\right)$ is the \emph{final particle
number inside a chosen comoving volume} during the radiation-dominated
era (a more thorough discussion of $N_{\mathrm{r}}\left(t_{rest}\right)$
will be held in Subsection \ref{sub:What-is-FCPN} and \ref{sub:CPN-meaning}).
The illustration for \eqref{eq:FPNo} is shown in Figure \ref{fig:FPNo}.
Thus, \eqref{eq:2.14}, \eqref{eq:EFF}, \eqref{eq:FPNo} and \eqref{eq:3.3}
provide
\begin{equation}
\left(\frac{dV}{d\phi}\right)_{t\geq t_{rest}}=-\sigma\left(t\right)\ell^{-1}N_{\mathrm{r}}^{\nicefrac{1}{3}}\left(t_{rest}\right)\label{eq:3.4}
\end{equation}
due to the balance between the force $-V^{'}\left(\phi\left(t\right)\right)$
and the EFF $\left(-\frac{4\Gamma\varepsilon_{\mathrm{r}}}{3\dot{\phi}}\right)$
when $t\geq t_{rest}$. In order to be consistent with the inference
of \eqref{eq:3.3}, $\sigma\left(t\right)$ must be a constant of
$\sigma\left(t_{rest}\right)$ (we call it the static reduced coefficient
of radiation creation, SRCRC), when $t\geq t_{rest}$.

Next, to fit observations, we must introduce a proper model of inflaton
into our discussion. We test result \eqref{eq:3.4} by employing \emph{the
classical chaotic model}, $V\left(\phi\right)=\frac{1}{2}m_{\phi}^{2}\phi^{2}$.
Since the remaining inflaton potential will become the energy density
of ECC, as $V\left(\phi\left(t_{rest}\right)\right)=\frac{\Lambda}{8\pi G}$,
$\Lambda$ and $N_{\mathrm{r}}\left(t_{rest}\right)$ will be found
in expression 
\begin{equation}
V^{'}\left(\phi\left(t_{rest}\right)\right)=\pm\sqrt{\frac{m_{\phi}^{2}\Lambda}{4\pi G}}=-\sigma\left(t_{rest}\right)\ell^{-1}N_{\mathrm{r}}^{\nicefrac{1}{3}}\left(t_{rest}\right).\label{eq:3.5}
\end{equation}
Here, the sign of ``$\pm$'' is dependent on the position of $\phi\left(t_{rest}\right)$.
Additionally, it should be noted that $\left\rangle \sigma\left(t_{rest}\right)\right\langle $
is opposite to $\left\rangle \phi\left(t_{rest}\right)\right\langle $.
Rewriting \eqref{eq:3.5} clearly, the exact $\Lambda$-CPN relationship
becomes
\begin{equation}
N_{\mathrm{r}}\left(t_{rest}\right)=\left(\frac{\ell^{2}m_{\phi}^{2}\Lambda}{4\pi G\sigma^{2}\left(t_{rest}\right)}\right)^{\nicefrac{3}{2}}.\label{eq:3.6}
\end{equation}

\subsection{Numerical results \label{apx:numerical}}

\subsubsection{The comoving entropy and the comoving particle number \label{sub:CE-CPN}}

Before we present the numerical result for the CPN and $\Lambda$,
we should first outline the relationship between comoving entropy
and the CPN. Nesteruk has come up with an applicable method and his
result is that the particle number of a decoupled species inside a
chosen adiabatic comoving volume is proportional to the decoupled
species' entropy as observed from the same volume (we call this comoving
entropy for short) \cite{Thermal_inflation-2}. According to the Gibbs
equation

\begin{equation}
T_{j}dS_{j}=d\left(\varepsilon_{j}v\right)+p_{j}dv,\label{eq:a.1.1}
\end{equation}
we can easily have 
\begin{equation}
T_{j}\frac{\dot{S}_{j}}{v}=\dot{\varepsilon}_{j}+\left(\varepsilon_{j}+p_{j}\right)\left(\frac{\dot{\mathbb{R}}}{\mathbb{R}}\right).\label{eq:a.1.2}
\end{equation}
Here, $v=\left(\mathbb{R}\left(t\right)\ell\right)^{3}$ is the chosen
comoving volume; $S_{j}\left(t\right)$ is the entropy for an arbitrary
species $j$ inside the chosen comoving volume; and $T_{j}$ is the
temperature defined by the density of species $j$. When the Universe
is creating the species $j$, $S_{j}\left(t\right)$ increases simultaneously.
Then, the entropy reaches a maximum constant when the creation of
the species has ceased and the species itself has decoupled from others.
In other words, the situation of $\dot{S}_{j}\left(t\geq t_{j,\,\mathrm{dec}}\right)=0$
means that there is no (net) energy current for the species $j$ between
arbitrary comoving volumes. It also means that no energy transfer
exists between species $j$ and other species inside the chosen comoving
volume.

Of course, $\underset{l}{\sum}S_{l}$ will finally become a maximum
constant to consist with the basic belief---the unique or adiabatic
Universe---when the creation of all of species has ceased, even if
some do not decouple from others. This is where the famous continuity
equation of the Standard Model of Cosmology comes in.

Now, we adopt the example of a decoupling-neutrino. The following
thermal equilibrium conditions are needed: 
\begin{eqnarray}
p_{\nu}\left(t\right) & = & \frac{\varepsilon_{\nu}\left(t\right)}{3},\label{eq:a.1.4}\\
\varepsilon_{\nu}\left(t\right) & = & \frac{\pi^{2}}{30}\cdot\frac{7}{8}\underset{\nu}{\sum}g_{\nu}T_{\nu}^{4}\left(t\right),\label{eq:a.1.5}\\
n_{\nu}\left(t\right) & = & \frac{\zeta\left(3\right)}{\pi^{2}}\cdot\frac{3}{4}\underset{\nu}{\sum}g_{\nu}T\nu^{3}\left(t\right),\label{eq:a.1.6}
\end{eqnarray}
at the cosmic time $t\gtrsim t_{\nu,\,\mathrm{dec}}$. Here $t_{\nu,\,\mathrm{dec}}\gtrsim\unit[0.2]{s}$
is the decoupling time of the neutrino; the decoupling temperature
is about $T_{\nu}\left(t\approx t_{\nu,\,\mathrm{dec}}\right)\simeq\unit[1.5]{MeV}$
\cite{Neu-1,early_universe,PFC}; $\underset{\nu}{\sum}g_{\nu}=6$
is the degree of freedom for neutrino species (3 flavors, 2 spin states
each). The Gibbs equation for the decoupling-neutrino becomes 
\begin{equation}
\left(T_{\nu}dS_{\nu}\right)_{t}\thickapprox4.2\cdot\left(T_{\nu}dN_{\nu}\right)_{t}\label{eq:a.1.7}
\end{equation}
in consideration of $N_{\nu}\left(t\right)=n_{\nu}\left(t\right)v\left(t\right)$.

However, we have no direct data for $S_{\nu}\left(t_{\mathrm{dec}}\right)$.
Fortunately, we can substitute $S_{\nu}\left(t_{\mathrm{now}}\right)$
for $S_{\nu}\left(t_{\mathrm{dec}}\right)$, and vice versa since
$\dot{S}_{\nu}\left(t\geq t_{\nu,\,\mathrm{dec}}\right)\approx0$
\footnote{Due to the possibility of a small mass, $\sum m_{\nu}<\unit[0.58]{eV}\:\left(\mathrm{for\,}w=-1\right)$
\cite{WMAP7}, the neutrino probably interacts weakly with gravity,
and it also looks like a non-relativistic particle at the present
moment. This is why the symbol of ``$\approx$'' is used. Regardless,
the neutrino number inside a large enough chosen comoving volume should
not change after its decoupling time.%
}. According to the estimate of \cite{entropy 1,entropy 2}, the entropy
of cosmic-background neutrinos inside presently observable Universe
is $S_{\mathrm{CBN}}^{obs}\left(t_{\mathrm{now}}\right)=s_{\mathrm{CBN}}\left(t_{\mathrm{now}}\right)\cdot v_{obs}\left(t_{\mathrm{now}}\right)=\left(5.16\pm0.14\right)\times10^{89}$.
This is because the present day value of the CBN entropy density is
\begin{eqnarray}
s_{\mathrm{CBN}}\left(t_{\mathrm{now}}\right) & = & \frac{2\pi^{2}}{45}\cdot\frac{7}{8}\underset{\nu}{\sum}g_{\nu}T_{\mathrm{CBN}}^{3}\left(t_{\mathrm{now}}\right)\nonumber \\
 & = & \unit[\left(1.411\pm0.014\right)\times10^{9}]{m^{-3}},\label{eq:entropy-density}
\end{eqnarray}
and the volume of today's observable Universe \cite{early_universe}
is
\begin{eqnarray}
v_{obs}\left(t_{\mathrm{now}}\right) & = & \frac{4}{3}\pi\left(\mathbb{R}\left(t_{\mathrm{now}}\right)\ell_{\mathrm{PH}}\right)^{3}\nonumber \\
 & = & \frac{4}{3}\pi\left(\mathbb{R}\left(t_{\mathrm{now}}\right)\int_{t_{\mathrm{Planck}}}^{t_{\mathrm{now}}}\frac{dt}{\mathbb{R}\left(t\right)}\right)^{3}\nonumber \\
 & = & \unit[\left(3.65\pm0.10\right)\times10^{80}]{m^{3}}.\label{eq:obs-volume}
\end{eqnarray}
(The present CBN temperature is $T_{\mathrm{CBN}}=\left(\nicefrac{4}{11}\right)^{\nicefrac{1}{3}}T_{\mathrm{CMB}}=\left(\nicefrac{4}{11}\right)^{\nicefrac{1}{3}}\cdot\left(\unit[2.725\pm0.002]{K}\right)$.
$\ell_{\mathrm{PH}}$ is the present particle horizon coordinate distance.)
Therefore, the neutrino number was about $N_{\nu}\left(t_{\mathrm{dec}}\right)\approx1.2\times10^{89}$
inside the comoving volume of the present particle horizon coordinate
distance ($v\left(t_{\nu,\,\mathrm{dec}}\right)\propto\left(\mathbb{R}\left(t_{\nu,\,\mathrm{dec}}\right)\ell_{\mathrm{PH}}\right)^{3}$)
at the decoupling time $t_{\nu,\,\mathrm{dec}}$. Of course, the neutrino
number inside the presently observable Universe ($v_{obs}\left(t_{\mathrm{now}}\right)\propto\left(\mathbb{R}\left(t_{\mathrm{now}}\right)\ell_{\mathrm{PH}}\right)^{3}$)
is still $N_{\nu}\left(t_{\mathrm{now}}\right)\approx1.2\times10^{89}$.

One point worthy of attention is the fact that result \eqref{eq:a.1.7}
can be applied to CMB and WIMP dark matter ($S\approx4N_{\mathrm{CMB\, or\, WDM}}$)
because they are the known decoupled species. Of course, this requires
the species of WDM to be a kind of fermion. And then we can employ
the same thought to estimate these particle number inside the comoving
volume for any chosen coordinate distance.

\subsubsection{What is the $N_{\mathrm{r}}\left(t_{rest}\right)$? \label{sub:What-is-FCPN}}

Now, we have to talk about the role and essence of $N_{\mathrm{r}}\left(t_{rest}\right)$.
In Section \ref{sec:Introduction}, the words of ``particle number''
has defined as the \emph{sum of lepton, baryon (quark) and gauge boson}
\emph{number}. However, we can also count the CPN by using the classification
of the species of particles (such as photon, electron, neutrino, quark,
and so on). Dependent on the consideration, the following statements
should be highlighted to facilitate further discussion:
\begin{enumerate}
\item According to Subsection \ref{sub:CE-CPN}, the CPN of an arbitrary
decoupled species $j$ ($N\left(t_{j,\,\mathrm{dec}}\right)$) is
proportional to its entropy ($S\left(t_{j,\,\mathrm{dec}}\right)$),
which can be obtained from a chosen comoving volume ($\left(\mathbb{R}\left(t_{j,\,\mathrm{dec}}\right)\ell\right)^{3}$)
at the time ($t_{j,\,\mathrm{dec}}$) when the species $j$ decouples
from others. Moreover, we can conclude that no net energy current
manifests between arbitrary comoving volumes inside the Universe,
not only because the decoupled species has stopped its interactions
with other species, but also due to the belief that our Universe is
unique or adiabatic. Therefore, the entropy of a decoupled species
inside a comoving volume should be a constant, even if the volume
is expanding. This is equivalent to the other fact: namely that the
CPN of an arbitrary decoupled species is also incontestably a constant.
\item Due to the spirit of the Bose-Einstein and Fermi-Dirac distributions,
we can employ the temperature to relate the (energy or number) density
of a \emph{relativistic species}. On the other hand, a relativistic
species' density can be employed to relate the temperature of the
species. (Here, the properties of $T_{\mathrm{b,\, f}}\left(t\right)\gg m_{\mathrm{b,\, f}}$
and $T_{\mathrm{b,\, f}}\left(t\right)\gg\mu_{\mathrm{b,\, f}}$ should
be satisfied.) Thus, when a species $j$ drops out of thermal equilibrium,
its temperature and density will evolve independently. This is the
meaning of the ``decoupling''. Since the system what we concern
is adiabatic, the comoving entropy of the species $j$ will be a constant
after its decoupling, namely, the CPN of the species $j$ will be
also invariant. For this reason, we absolutely can employ the time
of $t_{j,\,\mathrm{dec}}$ to represent the time while the species
$j$ is ceased to produce, and the constant CPN of the species $j$
can be shown as $N\left(t_{j,\,\mathrm{dec}}\right)$. From the viewpoint
of the particle physics, the decoupling of a species means that the
species will no longer interact with others. It is of course that
a decoupled species will not become another species.
\item Equations \eqref{eq:3.4} and \eqref{eq:3.5} remind us that the E(S)FF
is dependent on $N_{\mathrm{r}}\left(t_{\left(rest\right)}\right)$.
For this reason, any situation which can change this number will also
weaken or strengthen the EFF. Of course, the value of the cosmological
constant will be moved indirectly by the change of the CPN. (The situation
of $\Lambda=0$ is a very special case that will be discussed in Subsection
\ref{sub:probabilistic-L-CPN}.) Thus, in our scenario, a fixed ECC
implies that the Universe has a finite and constant CPN $N\left(t_{rest}\right)$.
However, according to the $\Lambda$-CPN relationship \eqref{eq:3.6},
an observable variation of $\Lambda$ needs a sufficient number fluctuation
to occur at the same cosmic time throughout the entire Universe. In
other words, if the comoving number fluctuation is \emph{not enough},
or the fluctuation is LOCAL (even if it is very big), the observable
variation of $\Lambda$ will not be discovered by us or any alien.
\end{enumerate}
Surely, the number $N_{\mathrm{r}}\left(t_{rest}\right)$ which appears
in this paper means the \emph{TOTAL} CPN, i.e., 
\begin{equation}
N_{\mathrm{r}}\left(t_{rest}\right)=\underset{j}{\sum}N\left(t_{j,\,\mathrm{dec}}\right)+\underset{l}{\sum}N_{l,\,\mathrm{int}}.\label{eq:CPN_1}
\end{equation}
As we know, some species which still have interactions with others
can be found in stars, the center of galaxies, our accelerators, nukes,
and even in light bulbs (photons only) etc. Due to the facts, we can
separate these species into the part of $\underset{l}{\sum}N_{l,\,\mathrm{int}}$.
It is now known that if we ignore the case of neutrino oscillation,
the lepton number and baryon (quark) number will be conserved after
the GUT (grand unified theory) phase. Put simply, $\underset{l}{\sum}N_{l,\,\mathrm{int}}$
can be regarded as the constant for a big enough comoving volume.
(Since, throughout the entire Universe, the energy density of the
photons created by stars and humankind is too small.)

According to the present data, the estimate of WIMP dark matter $S_{\mathrm{WDM}}^{obs}\left(t_{\mathrm{now}}\right)$
is close to $2\times10^{88\pm1}$ \cite{entropy 1}; CMB data supports
$S_{\mathrm{CMB}}^{obs}\left(t_{\mathrm{now}}\right)\approx5.4\times10^{89}$
\cite{entropy 1,entropy 2}. Including the CBN comoving number, these
provide a similar figure of $10^{89}$ inside the comoving volume
($\frac{4\pi}{3}\left(\mathbb{R}\left(t>t_{Ldec}\right)\ell_{\mathrm{PH}}\right)^{3}$,
$t_{Ldec}$ is the decoupling time for the last decoupled species).
On the other hand, the number density for presently \textquotedblleft{}observable\textquotedblright{}
baryon is about $n_{\mathrm{B}}\approx\unit[0.11]{m^{-3}}$ (Fukugita
et al., 1997 \cite{BND_1997}) or $n_{\mathrm{B}}\approx\unit[0.26]{m^{-3}}$
(Komatsu et al., 2010 \cite{WMAP7}), which means that the number
of baryons currently seen in cosmic gas, dust and stars inside the
Universe is about $10^{79}$. Therefore, we can conclude that $N_{\mathrm{r}}\left(t_{rest}\right)\approx10^{89\sim90}$
is the total particle number which is the sum of these major parts
of matter inside the comoving volume of $v_{\mathrm{PH}}\left(t\right)=\frac{4\pi}{3}\left(\mathbb{R}\left(t>t_{Ldec}\right)\ell_{\mathrm{PH}}\right)^{3}$.

\subsubsection{Numerical results of the $\Lambda$-CPN relationship \label{sub:Numerical-results}}

\begin{table}
\begin{centering}
\caption{\label{tab:N_vs_Lamda} Numerical results for $G\sigma^{2}\left(t_{rest}\right)$
vs. $N_{\mathrm{r}}\left(t_{rest}\right)$ vs. $S_{\mathrm{r}}\left(t\approx t_{Ldec}\right)$.
Inflaton mass is employed as $m_{\phi}=t_{\mathrm{Planck}}^{-1}\approx\unit[1.855\times10^{43}]{s^{-1}}$;
the value of ECC is $\Lambda\approx\unit[1.127\times10^{-35}]{s^{-2}}$,
which is dependent on the average present day value of the Hubble
rate, $H_{\mathrm{now}}\approx\unitfrac[70]{\unitfrac{km}{s}}{Mpc}$
\cite{acc:2011}. The comoving coordinate radius of the present particle
horizon is defined as the unit length: $\ell_{\mathrm{PH}}\equiv1$.
The volume of the presently observable Universe is $v_{obs}\left(t_{\mathrm{now}}\right)=\unit[3.65\times10^{80}]{m^{3}}$.}

\par\end{centering}

\begin{centering}
\begin{tabular}{cccc}
 &  &  & \tabularnewline
\hline 
\hline 
\enskip{}%
\begin{tabular}{c}
$\ell^{\:\dagger}$\tabularnewline
\end{tabular}\enskip{} & \enskip{}%
\begin{tabular}{c}
$G\sigma^{2}\left(t_{rest}\right)$\tabularnewline
\end{tabular}\enskip{} & \qquad{}%
\begin{tabular}{c}
$N_{\mathrm{r}}\left(t_{rest}\right)$\tabularnewline
\end{tabular}\qquad{} & \enskip{}%
\begin{tabular}{c}
$S_{\mathrm{r}}\left(t_{Ldec}\right)$\tabularnewline
\end{tabular}\enskip{}\tabularnewline
\hline 
\begin{tabular}{c}
$1$\tabularnewline
\end{tabular} & %
\begin{tabular}{c}
$1$\tabularnewline
\end{tabular} & %
\begin{tabular}{c}
$5.421\times10^{75}$\tabularnewline
\end{tabular} & %
\begin{tabular}{c}
$2.168\times10^{76}$\tabularnewline
\end{tabular}\tabularnewline
\begin{tabular}{c}
$1$\tabularnewline
\end{tabular} & %
\begin{tabular}{c}
$10^{-4}$\tabularnewline
\end{tabular} & %
\begin{tabular}{c}
$5.421\times10^{81}$\tabularnewline
\end{tabular} & %
\begin{tabular}{c}
$2.168\times10^{82}$\tabularnewline
\end{tabular}\tabularnewline
\begin{tabular}{c}
$1$\tabularnewline
\end{tabular} & %
\begin{tabular}{c}
$10^{-9.2}$\tabularnewline
\end{tabular} & %
\begin{tabular}{c}
$3.420\times10^{89}$\tabularnewline
\end{tabular} & %
\begin{tabular}{c}
$1.368\times10^{90}$\tabularnewline
\end{tabular}\tabularnewline
\hline 
\hline 
\begin{tabular}{c}
$0.01$\tabularnewline
\end{tabular} & %
\begin{tabular}{c}
$10^{-9.2}$\tabularnewline
\end{tabular} & %
\begin{tabular}{c}
$3.420\times10^{83}$\tabularnewline
\end{tabular} & %
\begin{tabular}{c}
$1.368\times10^{84}$\tabularnewline
\end{tabular}\tabularnewline
\end{tabular}
\par\end{centering}

\centering{}{\small $^{\dagger\;}\ell=1$ is the case of the present
particle horizon coordinate distance. }
\end{table}

Here, we provide numerical results in Table \ref{tab:N_vs_Lamda}.
These values are obtained by taking the value of inflaton mass and
the observational data of $\Lambda$ into \eqref{eq:3.6} and \eqref{eq:a.1.7}.

\newpage{}

\section{Probabilistic $\Lambda$ and the course of particle creation \label{sec:prob_L-PCC}}

\subsection{The behavior of $\sigma\left(t\right)$ \label{sub:RCRC-behavior}}

In Section \ref{sec:CPN-Lambda}, we define a RCRC, $\sigma\left(t\right)$,
for describing the course of particle creation. As in the equations
of \eqref{eq:EFF}, the requirement  $\left\rangle \sigma\left(t\right)\right\langle =\left\rangle \dot{\phi}\left(t\right)\right\langle $
guarantees that $\left\rangle \mathrm{EKFF}\right\langle $ is always
opposite to $\left\rangle \dot{\phi}\right\langle $. In addition,
according to \eqref{eq:3.5}, if we employ $V\left(\phi\right)=\frac{1}{2}m_{\phi}^{2}\phi^{2}$
to consider the case of an oscillating inflaton, $\left\rangle \sigma\left(t_{rest}\right)\right\langle =-\left\rangle \phi\left(t_{rest}\right)\right\langle $
is included to insure $\left\rangle \mathrm{ESFF}\right\langle $
against the direction of the final restoring force $\left\rangle -V^{'}\left(\phi\left(t_{rest}\right)\right)\right\langle $.

However, a sharp-eyed reader will find a discontinuous situation:
there exists a case whereby $\left\rangle \sigma\left(t\right)\right\langle =-\left\rangle \sigma\left(t_{rest}\right)\right\langle $
when $\phi$ approaches to the static point $\phi\left(t_{rest}\right)$
(the relationship of time is $t\rightarrow t_{rest}$). All of the
various behaviors when $t\rightarrow t_{rest}$ are listed in Table
\ref{tab:RCRC_vs_phi}.

\begin{table}
\begin{centering}
\caption{\label{tab:RCRC_vs_phi} This shows the relationship of signs (directions)
which occur at the time when $t\rightarrow t_{rest}$. Here we employ
the restoring force of $\phi$ as ``$-\frac{dV\left(\phi\left(t\right)\right)}{d\phi}=-m_{\phi}^{2}\phi\left(t\right)$''.}

\par\end{centering}

\centering{}%
\begin{tabular}{cccccccc}
 &  &  &  &  &  &  & \tabularnewline
\hline 
\hline 
$\left\rangle \sigma\left(t\right)\right\langle \;\:$ & $\left\rangle \mathbf{f}_{\phi k}\right\langle \;\:$ & $\left\rangle \sigma\left(t_{rest}\right)\right\langle \;\:$ & $\left\rangle \mathbf{f}_{\phi s}\right\langle \;\:$ & $\left\rangle \dot{\phi}\left(t\right)\right\langle \;\:$ & $\left\rangle -m_{\phi}^{2}\phi\left(t\right)\right\langle \;\:$ & $\left\rangle \phi\left(t_{rest}\right)\right\langle \;\:$ & $\left\rangle -m_{\phi}^{2}\phi\left(t_{rest}\right)\right\langle $\tabularnewline
\hline 
$+$ & $-$ & $+$ & $-$ & $+$ & $+$ & $-$ & $+$\tabularnewline
$-$ & $+$ & $-$ & $+$ & $-$ & $-$ & $+$ & $-$\tabularnewline
\hline 
$+$ & $-$ & $-$ & $+$ & $+$ & $-$ & $+$ & $-$\tabularnewline
$-$ & $+$ & $+$ & $-$ & $-$ & $+$ & $-$ & $+$\tabularnewline
\hline 
\hline 
 &  &  &  &  &  &  & \tabularnewline
\end{tabular}
\end{table}

The continuous situations between $\left\rangle \sigma\left(t\right)\right\langle $
and $\left\rangle \sigma\left(t_{rest}\right)\right\langle $ are
shown in the two columns above the central line of Table \ref{tab:RCRC_vs_phi}.
Meanwhile, the nethermost two columns are the discontinuous situations.
Actually, discontinuous situations do not only happen in the example
of the spring oscillating system from \cite{YCCHEN}; if we consider
the question carefully, we will see that they also occur when $\phi$
approaches each and every turning point. However, even though the
behavior of $\sigma\left(t\right)$ does not violate the laws of physics,
we remain uncomfortable because the discontinuation means that Nature
allows a $\sigma\left(t\right)$ which can suddenly change its sign
but keep its value. Until now, a reasonable solution to explain these
discontinuities has not been forthcoming. Fortunately, however, there
does exist an intriguing mechanism for illustrating them. This unexpected
result will be proposed in Subsection \ref{sub:RCRC-structure}.

\subsection{The structure of $\sigma\left(t\right)$ \label{sub:RCRC-structure}}

\subsubsection{The structure of $\sigma\left(t\right)$ and the new inflaton field equation \label{sub:new-infaton-FE}}

Now, we will try to provide illustrations for the discontinuity of
$\sigma\left(t\rightarrow t_{rest}\right)$ and the probabilistic
result of $\phi\left(t_{rest}\right)=0$. According to \eqref{eq:3.6},
$N_{\mathrm{r}}\ggg1$ and $\Lambda=0$ cannot occur simultaneously.
However, if we employ the inflaton model in the form of $V\left(\phi\right)=\frac{1}{2}m_{\phi}^{2}\phi^{2}$
and consider that a maximum ESFF exists in inflaton dynamics, the
discussion and conclusion of \cite{YCCHEN} indicates that, because
of (quantum) probability, it is possible for the final position of
$\phi\left(t_{rest}\right)$ inside the ``stagnant zone'' to become
zero. To account for this violation, a reasonable\textemdash{}if temporary\textemdash{}explanation
is that the SRCRC $\sigma\left(t_{rest}\right)$ will become zero
when $\phi\left(t_{rest}\right)$ is also zero. If this deduction
works, a possible structure of RCRC is
\begin{equation}
\sigma\left(t\right)=u\left(t\right)\phi\left(t\right)+\varpi\left(t\right)\dot{\phi}\left(t\right),\label{eq:4.1}
\end{equation}
where $u\left(t\right)$ and $\varpi\left(t\right)$ are two nonzero
and undetermined coefficients, and their units are $\left[u\left(t\right)\right]=\unit{s^{-1}}$
and $\left[\varpi\left(t\right)\right]=\unit{1}$. The first property
of $u\left(t\geq t_{rest}\right)=u\left(t_{rest}\right)$ must be
satisfied for a fixed $\sigma\left(t_{rest}\right)$ to be constructed,
but $\varpi\left(t\geq t_{rest}\right)$ does not necessarily have
to be a constant, since a zero value of $\dot{\phi}\left(t\geq t_{rest}\right)$
means that $\varpi\left(t\right)\dot{\phi}\left(t\right)$ contributes
nothing to $\sigma\left(t\geq t_{rest}\right)$. Then, a comparison
with Table \ref{tab:RCRC_vs_phi} and the analysis in Subsection \ref{sub:RCRC-behavior}
reveals that the conclusion 
\begin{equation}
\left\rangle u\left(t\right)\right\langle =-,\quad\left\rangle \varpi\left(t\right)\right\langle =+\label{eq:4.2}
\end{equation}
should be abided. 

Therefore, the discontinuity worry of $\sigma\left(t\rightarrow t_{r}\right)$
as listed in Table \ref{tab:RCRC_vs_phi} disappears due to the assumption
of \eqref{eq:4.1}. This result derives from two facts: the powerful
$\varpi\left(t_{\left\Vert \dot{\phi}\right\Vert >0}\right)\dot{\phi}\left(t_{\left\Vert \dot{\phi}\right\Vert >0}\right)$
makes $\sigma\left(t\right)$ become dependent on the (fast enough)
rolling $\dot{\phi}$; and $\sigma\left(t_{\dot{\phi}\rightarrow0}\right)$
is controlled by $-\left\Vert u\left(t_{\dot{\phi}\rightarrow0}\right)\right\Vert \phi\left(t_{\dot{\phi}\rightarrow0}\right)$.
Besides, any discontinuous course (including those that emerge when
$\phi$ approaches to every turning point) will have a situation of
$\sigma\left(t\lesssim\tau_{j}\right)=0$ (where $\left\{ \tau_{j}\right\} _{j=1,\,2,\,3,\,\cdots}$
is the set of times for these turning points). Thus, the EFF will
become zero when $t$ runs to $\tau_{j}$. An explanation of this
discovery is given in Subsection \ref{sub:Particle-annihilation}.
Readers can also review a toy example in Figure \ref{fig:EKFF}, which
will aid with picturing the EFF\textquoteright{}s behavior.

Followingly, \eqref{eq:2.14} will become \emph{the new inflaton field
equation}
\begin{equation}
\ddot{\phi}+\left(3H+\varpi\ell^{-1}N_{\mathrm{r}}^{\nicefrac{1}{3}}\right)\dot{\phi}+\left(V^{'}\left(\phi\right)-\left\Vert u\right\Vert \ell^{-1}\phi N_{\mathrm{r}}^{\nicefrac{1}{3}}\right)=0\label{eq:4.3}
\end{equation}
by introducing \eqref{eq:3.1} and \eqref{eq:4.1}. It should be mentioned
that $\ell$ is a constant chosen coordinate distance; $H$ and $N_{\mathrm{r}}$
must be functions of time; and $u\left(t\right)$ and $\varpi\left(t\right)$
are dependent on model selection. Clearly, \eqref{eq:4.3} is consistent
with the discussion of the dissipation term ($\Upsilon\dot{\phi}$,
which has been mentioned in footnote \ref{fn:field-equation-WI})
from the warm inflation scenario.

\subsubsection{The probabilistic $\Lambda$ and CPN \label{sub:probabilistic-L-CPN}}

If we employ the example of $\phi\left(t_{rest}\right)=\pm\sqrt{\frac{\Lambda}{4\pi Gm_{\phi}^{2}}}$
from the classical chaotic model for our Universe, the result  \eqref{eq:3.6}
will be transformed into
\begin{equation}
\Lambda=\frac{4\pi G}{m_{\phi}^{2}}\left(\left\Vert u_{\mathrm{our}}\left(t_{rest}\right)\right\Vert \ell^{-1}\phi\left(t_{rest}\right)N_{\mathrm{r}}^{\nicefrac{1}{3}}\left(t_{rest}\right)\right)^{2}\label{eq:4.4}
\end{equation}
according to the last term of \eqref{eq:4.3} when $\dot{\phi}\left(t_{rest}\right)=0$.
Here $u_{\mathrm{our}}\left(t_{rest}\right)$ means that the coefficient
is only for our Universe. From the relationship in \eqref{eq:4.4},
it looks as if the combination of $N_{\mathrm{r}}\ggg1$ and $\Lambda=0$
will possibly occur, provided that $\phi\left(t_{rest}\right)=0$. 

Now, two simpler questions require attention: What will happen to
the Universe if $\left\Vert u\left(t\right)\right\Vert =\left\Vert u_{\dagger}\right\Vert $
for $t\geq t_{begin}$? How about if $\left\Vert \sigma\left(t\right)\right\Vert =\left\Vert \sigma_{\dagger}\right\Vert $
for $t\geq t_{begin}$?

We first give answer to the second question. It is obvious that the
force relation of the oscillating $\phi$ is 
\begin{equation}
\left\Vert V^{'}\left(\phi\left(\tau_{j}\right)\right)\right\Vert \geq\left\Vert \mathbf{f}_{\phi k}\left(\tau_{j}\right)\right\Vert =\left\Vert \sigma_{\dagger}\right\Vert \ell^{-1}N_{\mathrm{r}}^{\nicefrac{1}{3}}\left(\tau_{j}\right),\quad\tau_{j}\leq t_{rest}\label{eq:4.5}
\end{equation}
at every turning point $\phi\left(\tau_{j}\right)$ ($\left\{ \tau_{j}\right\} _{j=1,\,2,\,3,\,\cdots}$
is the set of times for the turning points). Of course, $\phi\left(\tau_{j}\right)$
includes the last turning point---$\phi\left(t_{rest}\right)$. Employing
$V\left(\phi\right)=\frac{1}{2}m_{\phi}^{2}\phi^{2}$, \eqref{eq:4.5}
becomes
\begin{equation}
m_{\phi}^{2}\left\Vert \phi\left(t_{rest}\right)\right\Vert =\left\Vert \sigma_{\dagger}\right\Vert \ell^{-1}N_{\mathrm{r}}^{\nicefrac{1}{3}}\left(t_{rest}\right)\label{eq:4.6}
\end{equation}
when $N_{\mathrm{r}}\left(t_{rest}\right)$ becomes large enough at
$\tau_{k}=t_{rest}$. Dependent on equation \eqref{eq:4.6}, the Universe
has a \emph{nonzero exact ECC} of
\begin{equation}
\Lambda=\frac{4\pi G}{m_{\phi}^{2}}\left(\left\Vert \sigma_{\dagger}\right\Vert \ell^{-1}N_{\mathrm{r}}^{\nicefrac{1}{3}}\left(t_{rest}\right)\right)^{2}.\label{eq:4.7}
\end{equation}

Next, we will discuss the first question. Utilizing the model $V\left(\phi\right)=\frac{1}{2}m_{\phi}^{2}\phi^{2}$,
the fact of the force relationship is
\begin{equation}
\left\Vert V^{'}\left(\phi\left(\tau_{j}\right)\right)\right\Vert =m_{\phi}^{2}\left\Vert \phi\left(\tau_{j}\right)\right\Vert \geq\left\Vert u_{\dagger}\right\Vert \left\Vert \phi\left(\tau_{j}\right)\right\Vert \ell^{-1}N_{\mathrm{r}}^{\nicefrac{1}{3}}\left(\tau_{j}\right)\label{eq:4.8}
\end{equation}
for every turning point $\phi\left(\tau_{j}\right)$ (where $\tau_{j}\leq t_{rest}$).
Alternatively, the more clear relationship of \eqref{eq:4.8} is 
\begin{equation}
\left(m_{\phi}^{2}-\left\Vert u_{\dagger}\right\Vert \ell^{-1}N_{\mathrm{r}}^{\nicefrac{1}{3}}\left(\tau_{j}\right)\right)\left\Vert \phi\left(\tau_{j}\right)\right\Vert \geq0.\label{eq:4.9}
\end{equation}
Since the decreasing $\left\Vert \phi\left(t\right)\right\Vert $
provides a part of the energy to increase $N_{\mathrm{r}}\left(t\right)$,
the restoring force will subsequently be canceled out by the fixed
final EFF when $\tau_{j}\geq t_{rest}$, as
\begin{equation}
\left(m_{\phi}^{2}-\left\Vert u_{\dagger}\right\Vert \ell^{-1}N_{\mathrm{r}}^{\nicefrac{1}{3}}\left(t_{rest}\right)\right)\left\Vert \phi\left(t_{rest}\right)\right\Vert =0.\label{eq:4.10}
\end{equation}
It is obvious that the simplest conditions of balance are:
\begin{enumerate}
\item $N_{\mathrm{r}}\left(t_{rest}\right)=\left(\frac{m_{\phi}^{2}}{\left\Vert u_{\dagger}\right\Vert \ell^{-1}}\right)^{3}$
for any possible $\left\Vert \phi\left(t_{rest}\right)\right\Vert $
which is lower than the maximum value, $\left\Vert \phi\left(t_{rest}\right)\right\Vert _{\mathrm{MAX}}$.
Of course, this includes the situation of the minimum $\phi\left(t_{rest}\right)=0$.
Unfortunately, the value of $\left\Vert \phi\left(t_{rest}\right)\right\Vert _{\mathrm{MAX}}$
cannot be provided here unless we have sufficient information about
the energy transfer.
\item $\left\Vert \phi\left(t_{rest}\right)\right\Vert =0$ for any possible
value of $N_{\mathrm{r}}\left(t_{rest}\right)$. This means that  $N_{\mathrm{r}}\left(t_{rest}\right)<\left(\frac{m_{\phi}^{2}}{\left\Vert u_{\dagger}\right\Vert \ell^{-1}}\right)^{3}=N_{\mathrm{r}}\left(t_{rest}\right)_{\mathrm{MAX}}$
could occur. However, we also do not know the minimum value, $N_{\mathrm{r}}\left(t_{rest}\right)_{\mathrm{min}}$,
here %
\footnote{In addition, statement 3 of Subsection \ref{sub:What-is-FCPN} raises
a situation in which the amount of EFF and $\Lambda$ vary when CPN
is changed by some interactions. Now a proper explanation for the
special case of $\Lambda=0$ can be proposed: if the temporary static
position of $\phi\left(t_{TS}\right)=0$ is the minimum point of inflaton
potential, the number's change will not strengthen the corresponding
EFF, because of the two properties, $\sigma\left(t_{TS}\right)=u\left(t_{TS}\right)\phi\left(t_{TS}\right)$
and $\mathbf{f}_{TS}=-\sigma\left(t_{TS}\right)\ell^{-1}N_{\mathrm{r}}^{\nicefrac{1}{3}}\left(t_{TS}\right)$.
It follows that the EFF cannot be increased and is therefore unable
lift $\Lambda$ from zero. This is why the CPN is probabilistic, if
the ECC of a Universe is zero.%
}.
\end{enumerate}
According to the above two illustrations, $\phi\left(t_{rest}\right)$
and $N_{\mathrm{r}}\left(t_{rest}\right)$ can be concluded as the
probabilistic results that are restricted to the correspondingly necessary
regions. This is consistent with the expectation of the probabilistic
production of the cosmological constant \cite{YCCHEN}.

\subsubsection{The decrease of CPN and the reversed ``EKFF'' \label{sub:Particle-annihilation}}

Following the above discussion of RCRC, a very important discovery
should be mentioned. From the assumption of the dynamic RCRC in \eqref{eq:4.1},
it accords to have the situation of 
\begin{equation}
\left\rangle \sigma\left(t\right)\right\langle =-\left\rangle \dot{\phi}\left(t\right)\right\langle \label{eq:4.11}
\end{equation}
due to the contribution of a stronger ``$-\left\Vert u\left(t\right)\right\Vert \phi\left(t\right)$''
when the rolling $\phi$ \emph{approaches to every turning point}
of the $\phi$-system (while the relation of sign between $\phi$
and $\dot{\phi}$ is $\left\rangle \phi\left(t\right)\right\langle =\left\rangle \dot{\phi}\left(t\right)\right\langle $).
Such an unabandonable property will destroy the order of Postulate
B because we have 
\begin{equation}
\dot{N}_{\mathrm{r}}\left(t\right)=\frac{3}{4}\chi^{\nicefrac{4}{3}}\ell^{3}\left(-\left\Vert u\left(t\right)\right\Vert \phi\left(t\right)+\varpi\left(t\right)\dot{\phi}\left(t\right)\right)\mathbb{R}^{4}\left(t\right)\dot{\phi}\left(t\right)\label{eq:4.12}
\end{equation}
after combining \eqref{eq:3.2} and \eqref{eq:4.1}. Equation \eqref{eq:4.12}
depicts how the CPN will be decreased when \eqref{eq:4.11} (with
the stronger ``$-\left\Vert u\left(t\right)\right\Vert \phi\left(t\right)$'')
happens. Fortunately, all of the ideas outlined in this paper are
feasible, even if the temporary violation of Postulate B must be allowed
in our scenario. Moreover, if the guess of RCRC, \eqref{eq:4.1},
can be extended to any model of inflaton potential, the corresponding
Universe will also have some stages that the CPN will be decreased
when $\left(-\left\Vert u\left(t\right)\right\Vert \phi\left(t\right)+\varpi\left(t\right)\dot{\phi}\left(t\right)\right)\dot{\phi}\left(t\right)<0$.
In Appendix \ref{apx:PC-example}, we offer a toy example to illustratively
demonstrate the course of particle creation based on the assumption
from \eqref{eq:4.1}.

A careful analysis of the situations corresponding to \eqref{eq:4.11}
reveals that the direction of EKFF ($\mathbf{f}_{\phi k}=-\sigma\left(t\right)\ell^{-1}N_{\mathrm{r}}^{\nicefrac{1}{3}}\left(t\right)$)
will change at the stage of CPN decrease. It also asserts that the
$\phi$-system can have several ``frictionless'' situations when
$\phi\rightarrow\phi\left(\tau_{j}\right)$. Since the reversed \textquotedblleft{}EKFF\textquotedblright{}
weakens the restoring force  $-V^{'}\left(\phi\left(t\right)\right)$
when the rolling $\phi$ approaches to the turning point, it helps
$\dot{\phi}$ to keep running longer. Subsequently, particles will
begin to be created when $\phi$ starts to leave the turning point.
Again, these processes are presented in an easily digestible form
in the toy example from Appendix \ref{apx:PC-example}.

\subsection{The meaning of $N_{\mathrm{r}}\left(t\right)$ \label{sub:CPN-meaning}}

Even though we have illustrated the FCPN by employing the classification
of species in Subsection \ref{sub:What-is-FCPN}, in this subsection,
we should have a deeper discussion about the CPN from the point of
the quantum numbers---lepton and baryon (quark) number. Especially,
the discovery of the decrease of CPN can be found in our scenario.
According to the definition at the beginning of this article, the
name of particle number means the sum of lepton, baryon (quark) and
gauge boson number. In addition, the standard model of particle physics
tells us that the numbers of leptons and baryons will be conserved
after the GUT phase transition. Therefore, except the primordial cosmic
photons, the CPN can be determined as a constant after the GUT phase.
(The gluon number is similar to baryon number since gluons will combine
with quarks to become hadrons. Meanwhile, due to the electrical neutrality
of our Universe, the number of electrons is also close to the number
of quarks.) In other words, the lepton and baryon (quark) numbers
will not change if the annihilations occur. A question arises naturally:
what is the meaning of the decrease of the CPN indicated in the previous
sections?

\begin{table}
\caption{\label{tab:creation-conditions}The creation conditions of particles
and anti-particles. The sign of $\pm$ denote the sign for the value
of $\dot{\phi}\left(t\right)\cdot\dot{\phi}\left(t\right)$ and $-\phi\left(t\right)\cdot\dot{\phi}\left(t\right)$.
In addition, $P$ represents the creation of particle, $AP$ represents
the creation of anti-particle. }

\centering{}%
\begin{tabular}{c|c|c}
\multicolumn{1}{c}{} & \multicolumn{1}{c}{} & \tabularnewline
\hline 
\hline 
 & $\quad\dot{\phi}\left(t\right)\quad$ & $-\phi\left(t\right)$\tabularnewline
\hline 
$\quad\dot{\phi}\left(t\right)\quad$ & $+$ & $\quad+\qquad-\quad$\tabularnewline
\hline 
creation & $P$ & $\quad P\qquad AP\quad$\tabularnewline
\hline 
\hline 
\multicolumn{1}{c}{} & \multicolumn{1}{c}{} & \tabularnewline
\end{tabular}
\end{table}

In our opinion, the decrease presented in equation \eqref{eq:4.12}
can be illustrated by using the thought of anti-particle creation.
To image the process of particle creation, if $\dot{N}_{\mathrm{r}}\left(t\right)>0$
represents the phenomenon that the creation rate of particles is larger
than the rate of anti-particles, the course that the creation rate
of anti-particles is more than the rate of particles can be indubitably
denoted by $\dot{N}_{\mathrm{r}}\left(t\right)<0$. Moreover, if we
regard neutrino as the Majorana particle, the numbers of neutrinos
and cosmic photons will increase when their anti-particles create.
For this reason, it seems that our discovery of equation \eqref{eq:4.12}
can provide the illustration in phenomenon for the fact that the numbers
of baryons and electrons are much less than the numbers of photons
and neutrinos.

Dependent on \eqref{eq:4.12}, the evolution equation of particle
number, Table \ref{tab:creation-conditions} can be made to indicate
the conditions for particle and anti-particle creation. Clearly, the
absolute positive term $\dot{\phi}\left(t\right)\cdot\dot{\phi}\left(t\right)$
creates particles throughout the creation course for any arbitrary
situation of inflaton's motion. However, the creation of anti-particles
is controlled by the term of $-\dot{\phi}\left(t\right)\cdot\phi\left(t\right)$
only if the situation of $\left\rangle \dot{\phi}\left(t\right)\right\langle =\left\rangle \phi\left(t\right)\right\langle $
occurs. Moreover, it can easily discovered that the term of $\left\Vert u\left(t\right)\right\Vert \mathbb{R}^{4}\left(t\right)$
plays as the major role to define the amount of anti-particles. Therefore,
from the research of phenomenon, the facts about anti-particle for
its creation and less amount can be naturally obtained. In other words,
it is possible to depict the evolution of the CPN and the scale factor
during the creation epoch, if we can determine the terms of $u\left(t\right)$
and $\varpi\left(t\right)$.

Now, the conclusions of the CPN can be proposed from the point of
the quantum number: 1. lepton, baryon (quark) and primordial gauge
boson numbers will be ceased to create during the GUT phase transition;
2. the anti-particles will be created before the GUT phase transition
to reduce the particle number; 3. the number of the Majorana particles
will be increased when their anti-particles are created from the vacuum
or the interactions of other species; 4. the final total CPN will
become a constant after the stage of annihilation.

\section{Conclusions \label{sec:Conclusion}}

To our surprise, the $\Lambda$-CPN relationship is very simple. According
to equations \eqref{eq:3.1}, \eqref{eq:3.2} and \eqref{eq:3.6}
, the following statements may be possible:
\begin{enumerate}
\item The guess of $\dot{N}_{\mathrm{r}}\left(t\right)=\alpha\left(t\right)\ell^{3}\mathbb{R}^{4}\left(t\right)\cdot\dot{\phi}\left(t\right)$
is correct. Additionally, when $t\geq t_{rest}$, CRC should become
a constant of $\alpha\left(t_{rest}\right)$ to provide inflaton dynamics
with a nonzero effective static frictional force, even if particles
are no longer created.
\item According to \eqref{eq:3.2}, a static inflaton verifies why we have
never observed any spontaneously created particle. In other words,
the running inflaton controls the creation switch.
\end{enumerate}
Moreover, through the analysis of RCRC, we obtain the following conclusions:
\begin{itemize}
\item The final results of the ECC $\Lambda$ and the CPN $N\left(t_{rest}\right)$
are probabilistic. Their values will fall into their own ranges. For
$\Lambda$, the lowest minimum that can be allowed is zero, but the
upper limit can not be provide at the present stage. In addition,
the range of the CPN is $N\left(t_{rest}\right)_{\mathrm{min}}\leq N\left(t_{rest}\right)\leq\left(\frac{m_{\phi}^{2}}{\left\Vert u_{\dagger}\right\Vert \ell^{-1}}\right)^{3}$.
However, the minimum value of the CPN can not be determined.
\item We discover a new inflaton field equation of \eqref{eq:4.3} which
contains parameters of $N_{\mathrm{r}}\left(t\right)$ and ``particle
creation coefficients''. This equation shows that the EFF includes
a dissipation term, a kind of damping force of inflaton, and a friction
term which is dependent on whether the position of an inflaton can
provide a static EFF to keep the (tiny) relic of inflaton potential
remaining. 
\item Furthermore, since an inflaton still rolls after the end of inflation,
a rolling inflaton which \textbf{\emph{approaches to every turning
point}}\emph{ }of the $\phi$-system will lead the \textquotedblleft{}EKFF\textquotedblright{}
to reverse. In phenomenon, such a process makes the decrease of CPN
occur. This fact concludes that Postulate A is activated but Postulate
B destroyed during the last phase of inflation. Equation \eqref{eq:a.2.1}
and the figures of Appendix \ref{apx:PC-example} clearly show the
evolution of the CPN. To this result, it possibly relates with the
``course of antiparticle creation''. Due to the discovery of equation
\eqref{eq:4.12}, the problem of the short of baryon number can be
reasonably illustrated in phenomenon.
\end{itemize}
Comparing the estimates proposed by \cite{entropy 1,entropy 2,BND_1997,WMAP7}
with our numerical results from Table \ref{tab:N_vs_Lamda}, the evidence
strongly indicates that \textbf{\emph{the ECC is highly supported
by contributions from the CPN of the decoupled species of CBN, CMB
and WDM}}. Moreover, if the species of WDM is a kind of lepton or
baryon, it is very possible that the CMB photon is the last decoupled
species. Therefore, according to statement 3 of Subsection \ref{sub:What-is-FCPN}
and conclusion 4 of Subsection \ref{sub:CPN-meaning}, the CPN will
become a constant when the course of $e^{\pm}$ annihilation ends.
Moving forward, \textbf{\emph{the ECC that we observe today has been
stable since our Universe was a little more than $\mathbf{10^{0}}$
seconds old}}.

As for the result of the $10^{89}$ comoving particles which appear
in the presently observable Universe, we should employ $G\sigma^{2}\left(t_{rest}\right)\approx10^{-9}$
to connect the relationship between ECC and $N_{\mathrm{PH}}$ ($N_{\mathrm{PH}}$
is the particle number inside the comoving volume ($\frac{4\pi}{3}\left(\mathbb{R}\left(t>t_{Ldec}\right)\ell_{\mathrm{PH}}\right)^{3}$)).
Is there any mystery of $G\sigma^{2}\left(t_{rest}\right)\approx10^{-9}$?
We will continue our research to find out the answer of the question.

\begin{acknowledgments}
I sincerely thank my best friends Dan and Dr. Tsung-Che Liu for their
great suggestions, discussions, support and help. I also appreciate
the important talks that I have shared with Dr. Jiwoo Nam and Dr.
Shi Pu.
\end{acknowledgments}

\appendix

\section{The course of particle creation: A toy case \label{apx:PC-example}}

Here, we would like to offer an example to show the process of particle
creation. This is dependent on the assumption of \eqref{eq:4.1} for
RCRC. Again, it should be emphasized that the \textbf{following example
is a toy case designed to facilitate a smooth understanding of the
effect of \eqref{eq:4.1}}. We employ the CPN evolution equation
\begin{eqnarray}
N_{\mathrm{r}}\left(t\right) & = & N_{\mathrm{r}}\left(t_{begin}\right)+\label{eq:a.2.1}\\
 &  & \frac{3}{4}\chi^{\nicefrac{4}{3}}\ell^{3}\int_{t_{begin}}^{t}\left(-\left\Vert u\left(t\right)\right\Vert \phi\left(t\right)+\varpi\left(t\right)\dot{\phi}\left(t\right)\right)\mathbb{R}^{4}\left(t\right)\dot{\phi}\left(t\right)dt\nonumber 
\end{eqnarray}
from integrating equation \eqref{eq:3.2}. For drawing pictures,
the conditions 
\[
t_{begin}=1,\;\: N_{\mathrm{r}}\left(1\right)=25,\;\:\chi=0.8144,\;\:\ell=\ell_{\mathrm{PH}}=1,
\]
\[
\left\Vert u\left(t\right)\right\Vert =\varpi\left(t\right)=10^{10},
\]
are necessary. Besides, we order the scale factor and scalar field
as 
\[
\mathbb{R}\left(t\right)=\sqrt{t},
\]
\[
\dot{\phi}\left(t\right)=-\exp\left(-\frac{t-1}{4\pi}\right)\sin\left(\frac{t-1}{\sqrt{12\pi}}\right),\;\:\phi\left(1\right)\thickapprox4.9567.
\]
\begin{figure}[H]
\begin{centering}
\caption{\label{fig:phi-evolution}The evolution of scalar field $\phi\left(t\right)$.}

\par\end{centering}

\begin{centering}
\includegraphics[scale=0.63]{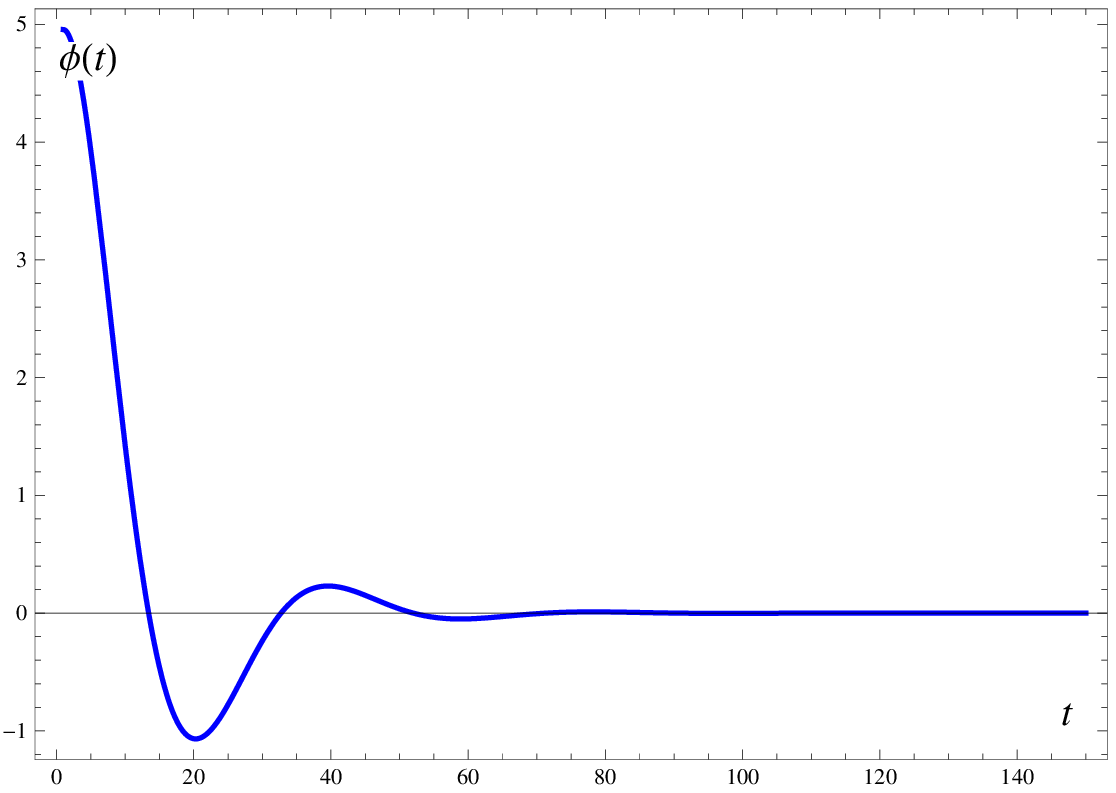}$\quad$\includegraphics[scale=0.63]{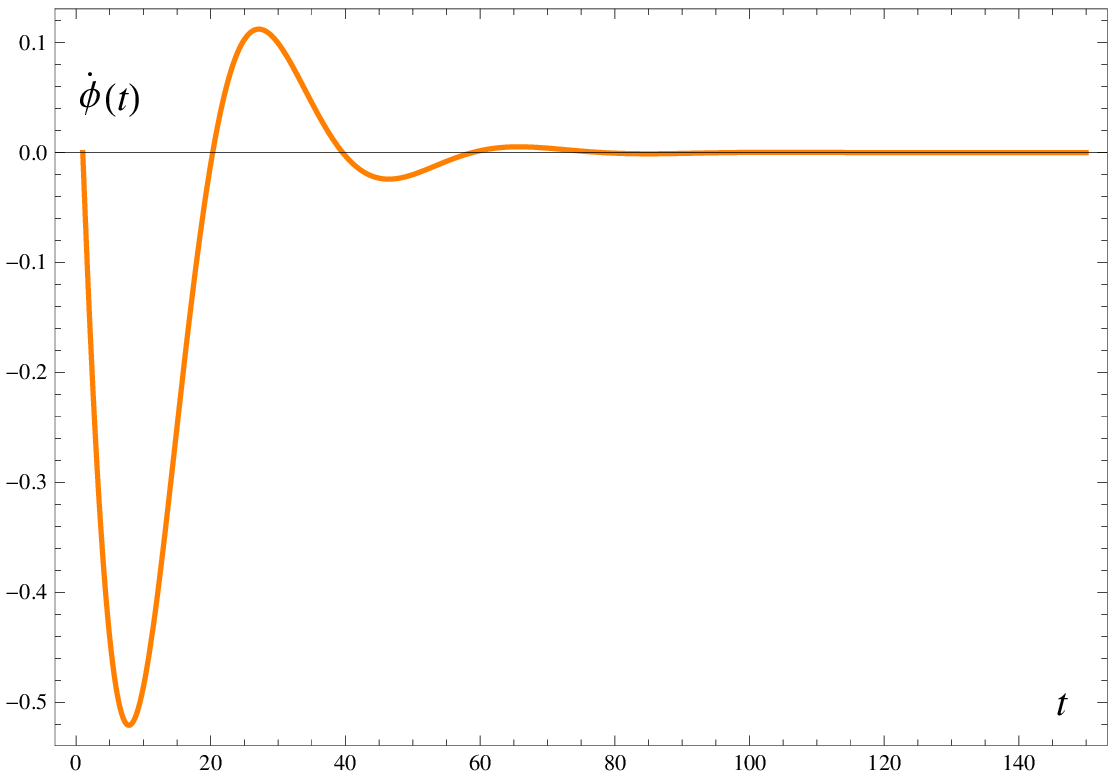}
\par\end{centering}

\begin{centering}
$\vphantom{}$
\par\end{centering}

\begin{centering}
\caption{\label{fig:EKFF}The evolution of effective friction.}

\par\end{centering}

\begin{centering}
\includegraphics[scale=0.63]{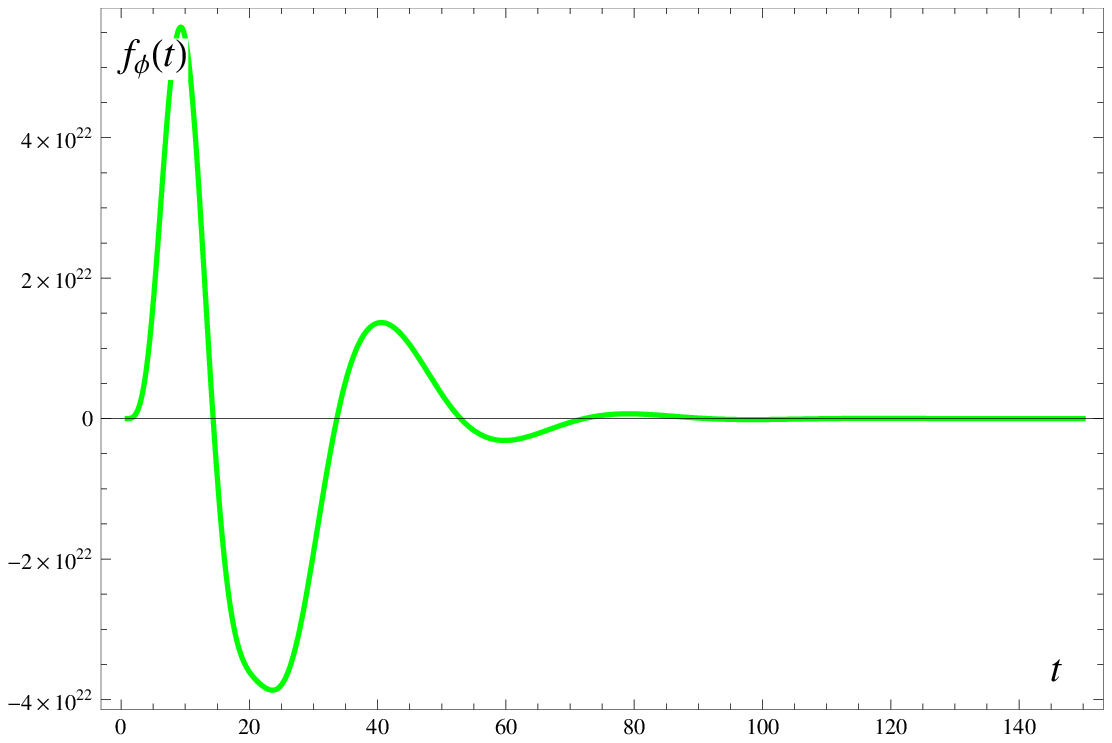}
\par\end{centering}

\begin{centering}
$\vphantom{}$
\par\end{centering}

\begin{centering}
\caption{\label{fig:No.-evolution}The evolution of particle number inside
a chosen comoving volume.}

\par\end{centering}

\centering{}\includegraphics[scale=0.62]{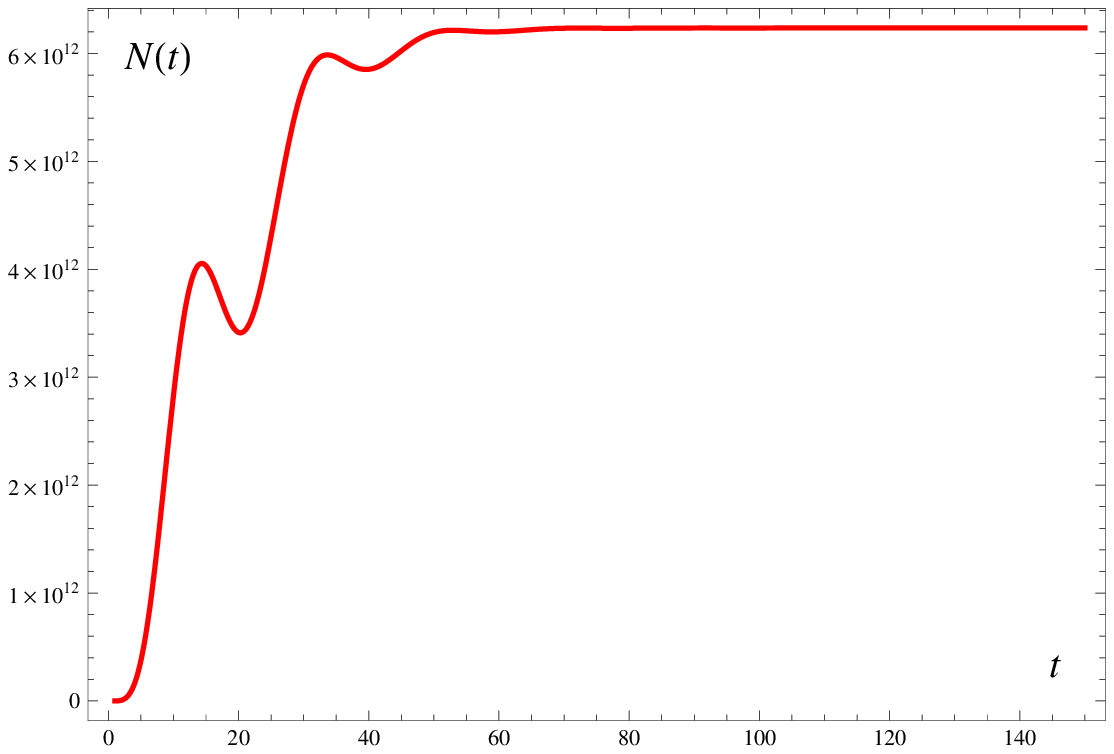}$\quad$\includegraphics[scale=0.65]{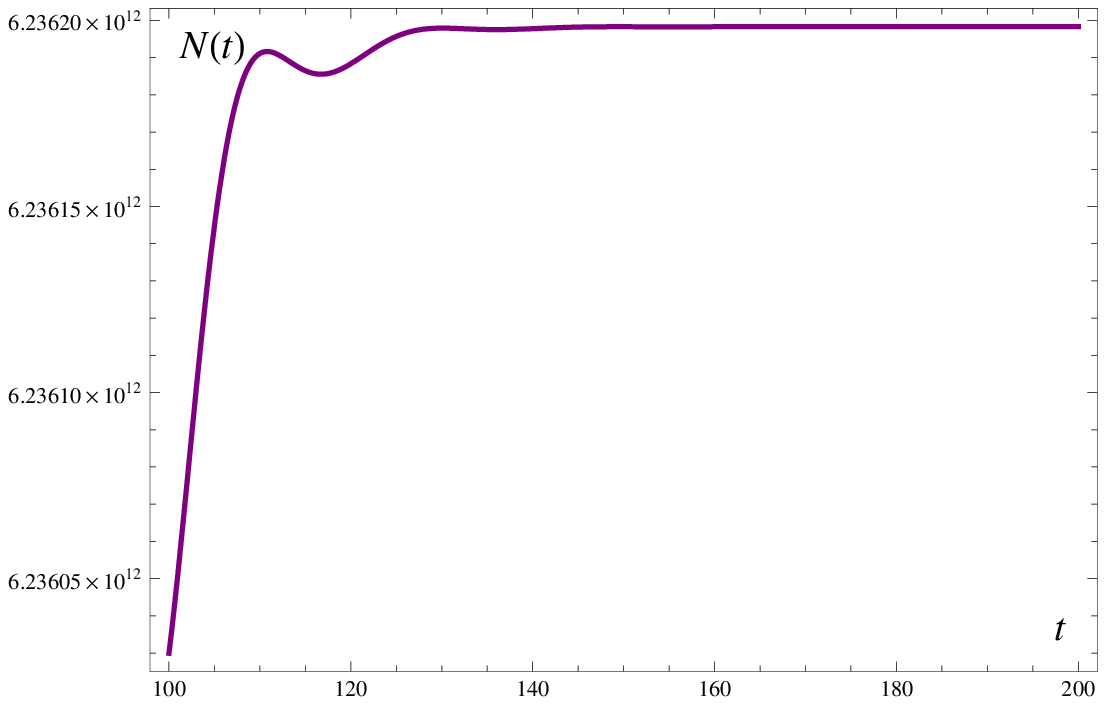}
\end{figure}
The course of particle creation can be clearly observed from Figure
\ref{fig:No.-evolution}. The EKFF reverses its direction from ``$+$''
to ``$-$'' for the first time at $t\approx15$, while $\left\rangle \dot{\phi}\left(t<20\right)\right\langle =-$.
Additionally, the result that the EKFF and $\dot{\phi}$ have the
same direction at the same time i.e., $\left\rangle \mathrm{f}_{\phi k}\left(15<t<20\right)\right\langle =\left\rangle \dot{\phi}\left(15<t<20\right)\right\langle =-$,
naturally follows. This causes the first instance of the decrease
of CPN to occur.

\newpage{}

\begin{table}
\caption{\label{tab:symbol-definition} AN ILLUSTRATION OF SYMBOLS THAT APPEAR
IN THIS PAPER}

\centering{}{\footnotesize }%
\begin{tabular}{cccc}
 &  &  & \tabularnewline
\hline 
\hline 
{\scriptsize }%
\begin{tabular}{c}
{\scriptsize SYMBOL}\tabularnewline
\hline 
\emph{\scriptsize NAME}\tabularnewline
\end{tabular} & {\scriptsize $\mathrm{UNIT^{\,\star}}$} & {\scriptsize $\mathrm{SIGN^{\:\ddagger}}$} & {\scriptsize MEANING}\tabularnewline
\hline 
\hline 
{\scriptsize }%
\begin{tabular}{c}
{\scriptsize $t$}\tabularnewline
\hline 
\emph{\scriptsize cosmic time}\tabularnewline
\end{tabular} & {\scriptsize $\unit{s}$} & {\scriptsize $+$} & {\scriptsize }%
\begin{tabular}{c}
{\scriptsize $t_{begin}$}\tabularnewline
\hline 
{\scriptsize The time when inflation begins.}\tabularnewline
\end{tabular}\tabularnewline
\hline 
{\scriptsize }%
\begin{tabular}{c}
{\scriptsize $G$}\tabularnewline
\hline 
\emph{\scriptsize Newton's gravitational constant}\tabularnewline
\end{tabular} & {\scriptsize $\unit{s^{2}}$} & {\scriptsize $+$} & \tabularnewline
\hline 
{\scriptsize }%
\begin{tabular}{c}
{\scriptsize $\ell$}\tabularnewline
\hline 
\emph{\scriptsize comoving coordinate distance}\tabularnewline
\end{tabular} & {\scriptsize $\unit{s}$} & {\scriptsize $+$} & {\scriptsize }%
\begin{tabular}{c}
{\scriptsize This is chosen and fixed.}\tabularnewline
\hline 
{\scriptsize For more information see footnote \ref{fn:coordinate-distance}.}\tabularnewline
\end{tabular}\tabularnewline
\hline 
{\scriptsize }%
\begin{tabular}{c}
{\scriptsize $\mathbb{R}\left(t\right)$}\tabularnewline
\hline 
\emph{\scriptsize spatial scale factor}\tabularnewline
\end{tabular} & {\scriptsize $1$} & {\scriptsize $+$} & {\scriptsize $\mathbb{R}\left(t_{\mathrm{now}}\right)\equiv1$.}\tabularnewline
\hline 
{\scriptsize }%
\begin{tabular}{c}
{\scriptsize $\mathrm{v}$}\tabularnewline
\hline 
\emph{\scriptsize comoving coordinate volume}\tabularnewline
\end{tabular} & {\scriptsize $\unit{s^{3}}$} & {\scriptsize $+$} & {\scriptsize $\mathrm{v}=\ell^{3}$}\tabularnewline
\hline 
{\scriptsize }%
\begin{tabular}{c}
{\scriptsize $v\left(t\right)$}\tabularnewline
\hline 
\emph{\scriptsize comoving volume}\tabularnewline
\end{tabular} & {\scriptsize $\unit{s^{3}}$} & {\scriptsize $+$} & {\scriptsize }%
\begin{tabular}{c}
{\scriptsize $v\left(t\right)=\left(\mathbb{R}\left(t\right)\ell\right)^{3}$}\tabularnewline
\hline 
{\scriptsize The proper volume.}\tabularnewline
\end{tabular}\tabularnewline
\hline 
{\scriptsize }%
\begin{tabular}{c}
{\scriptsize $N\left(t\right)$}\tabularnewline
\hline 
\emph{\scriptsize comoving particle number}\tabularnewline
\end{tabular} & {\scriptsize $1$} & {\scriptsize $+$} & {\scriptsize Amount of particles inside a comoving volume. }\tabularnewline
\hline 
{\scriptsize }%
\begin{tabular}{c}
{\scriptsize $n\left(t\right)$}\tabularnewline
\hline 
\emph{\scriptsize number density}\tabularnewline
\end{tabular} & {\scriptsize $\unit{s^{-3}}$} & {\scriptsize $+$} & {\scriptsize $n\left(t\right)\equiv\frac{N\left(t\right)}{v\left(t\right)}$}\tabularnewline
\hline 
{\scriptsize }%
\begin{tabular}{c}
{\scriptsize $\varepsilon\left(t\right)$}\tabularnewline
\hline 
\emph{\scriptsize energy density}\tabularnewline
\end{tabular} & {\scriptsize $\unit{s^{-4}}$} & {\scriptsize $+$} & {\scriptsize }%
\begin{tabular}{c}
{\scriptsize $\varepsilon\left(t\right)\equiv\frac{E\left(t\right)}{v\left(t\right)}$;}\tabularnewline
\hline 
{\scriptsize $E\left(t\right)$ is the comoving energy.}\tabularnewline
\end{tabular}\tabularnewline
\hline 
{\scriptsize }%
\begin{tabular}{c}
{\scriptsize $p\left(t\right)$}\tabularnewline
\hline 
\emph{\scriptsize pressure}\tabularnewline
\end{tabular} & {\scriptsize $\unit{s^{-4}}$} & {\scriptsize $\pm$} & {\scriptsize }%
\begin{tabular}{c}
{\scriptsize $p_{\mathrm{DE}}<-\frac{\varepsilon_{\mathrm{DE}}}{3}$
for dark energy;}\tabularnewline
\hline 
{\scriptsize $p_{\mathrm{r}}=\frac{\varepsilon_{\mathrm{r}}}{3}$
for radiation.}\tabularnewline
\end{tabular}\tabularnewline
\hline 
{\scriptsize }%
\begin{tabular}{c}
{\scriptsize $H\left(t\right)$}\tabularnewline
\hline 
\emph{\scriptsize Hubble rate}\tabularnewline
\end{tabular} & {\scriptsize $\unit{s^{-1}}$} & {\scriptsize $\pm$} & {\scriptsize $H\left(t\right)\equiv\frac{\dot{\mathbb{R}}\left(t\right)}{\mathbb{R}\left(t\right)}$}\tabularnewline
\hline 
\end{tabular}
\end{table}

\newpage{}

\begin{table}
\begin{centering}
\begin{tabular}{cccc}
 &  &  & \tabularnewline
\hline 
{\scriptsize }%
\begin{tabular}{c}
{\scriptsize $\phi\left(t\right)$}\tabularnewline
\hline 
\emph{\scriptsize inflaton}\tabularnewline
\end{tabular} & {\scriptsize $\unit{s^{-1}}$} & {\scriptsize $\pm$} & {\scriptsize }%
\begin{tabular}{c}
{\scriptsize A real scalar field;}\tabularnewline
\hline 
{\scriptsize $\phi$ comes to rest at $t\geqslant t_{rest}$.}\tabularnewline
\end{tabular}\tabularnewline
\hline 
{\scriptsize }%
\begin{tabular}{c}
{\scriptsize $m_{\phi}$}\tabularnewline
\hline 
\emph{\scriptsize inflaton mass}\tabularnewline
\end{tabular} & {\scriptsize $\unit{s^{-1}}$} & {\scriptsize $+$} & \tabularnewline
\hline 
{\scriptsize }%
\begin{tabular}{c}
{\scriptsize $V\left(\phi\right)$}\tabularnewline
\hline 
\emph{\scriptsize inflaton potential}\tabularnewline
\end{tabular} & {\scriptsize $\unit{s^{-4}}$} & {\scriptsize $+$} & {\scriptsize }%
\begin{tabular}{c}
{\scriptsize $V\left(\phi\left(t_{rest}\right)\right)$}\tabularnewline
\hline 
{\scriptsize The remaining inflaton potential.}\tabularnewline
\end{tabular}\tabularnewline
\hline 
{\scriptsize }%
\begin{tabular}{c}
{\scriptsize $\Lambda$}\tabularnewline
\hline 
\emph{\scriptsize effective cosmological constant}\tabularnewline
\end{tabular} & {\scriptsize $\unit{s^{-2}}$} & {\scriptsize $+$} & {\scriptsize $\Lambda=8\pi GV\left(\phi\left(t_{rest}\right)\right)$.}\tabularnewline
\hline 
{\scriptsize }%
\begin{tabular}{c}
{\scriptsize $T\left(t\right)$}\tabularnewline
\hline 
\emph{\scriptsize temperature}\tabularnewline
\end{tabular} & {\scriptsize $\unit{s^{-1}}$} & {\scriptsize $+$} & \tabularnewline
\hline 
{\scriptsize }%
\begin{tabular}{c}
{\scriptsize $\mathrm{f}$}\tabularnewline
\hline 
\emph{\scriptsize effective friction}\tabularnewline
\end{tabular} & {\scriptsize $\unit{s^{-3}}$} & {\scriptsize $\pm$} & \tabularnewline
\hline 
{\scriptsize }%
\begin{tabular}{c}
{\scriptsize $\alpha\left(t\right)$}\tabularnewline
\hline 
\emph{\scriptsize coefficient of radiation creation}\tabularnewline
\end{tabular} & {\scriptsize $\unit{s^{-2}}$} & {\scriptsize $\pm$} & {\scriptsize $\alpha\left(t_{rest}\right)$ is a constant called the
SCRC.}\tabularnewline
\hline 
{\scriptsize }%
\begin{tabular}{c}
{\scriptsize $\sigma\left(t\right)$}\tabularnewline
\hline 
\emph{\scriptsize reduced coefficient of radiation creation}\tabularnewline
\end{tabular} & {\scriptsize $\unit{s^{-2}}$} & {\scriptsize $\pm$} & {\scriptsize $\sigma\left(t_{rest}\right)$ is a constant called the
SRCRC.}\tabularnewline
\hline 
{\scriptsize }%
\begin{tabular}{c}
{\scriptsize $\chi$}\tabularnewline
\hline 
\emph{\scriptsize particle number integral constant}\tabularnewline
\end{tabular} & {\scriptsize $1$} & {\scriptsize $+$} & {\scriptsize }%
\begin{tabular}{c}
{\scriptsize $\chi\approx0.8144$}\tabularnewline
\hline 
{\scriptsize Dependent on the thermal equilibrium condition.}\tabularnewline
\end{tabular}\tabularnewline
\hline 
{\scriptsize }%
\begin{tabular}{c}
{\scriptsize $S$}\tabularnewline
\hline 
\emph{\scriptsize comoving entropy}\tabularnewline
\end{tabular} & {\scriptsize $1$} & {\scriptsize $+$} & {\scriptsize It is estimated from a chosen comoving volume.}\tabularnewline
\hline 
\hline 
 &  &  & \tabularnewline
\end{tabular}
\par\end{centering}

\begin{centering}
{\scriptsize $^{\star}$ It is due to the unit choice that we adopt
the Natural units: $c=k_{\mathrm{B}}=\hbar=1$.}
\par\end{centering}{\scriptsize \par}

\begin{centering}
{\scriptsize $^{\ddagger}$ We denote the sign or direction of $\Omega$
by using $\left\rangle \Omega\right\langle $.}
\par\end{centering}{\scriptsize \par}

\end{table}


\begin{thebibliography}{10}
\bibitem{Einstein}A. Einstein, \emph{Kosmologische Betrachtungen
zur allgemeinen Relativit{\"a}tstheorie}, Sitzungsberichte der esK{\"o}niglich
Preu{\ss}ischen Akademie der Wissenschaften (Berlin), (1917) 142
{[}\href{http://nausikaa2.mpiwg-berlin.mpg.de/cgi-bin/toc/toc.x.cgi?dir=S250UZ0K&step=thumb}{ECHO}{]}.

\bibitem{Friedman}A. Friedman, \emph{{\"U}ber die Kr{\"u}mmung
des Raumes}, \href{http://www.springerlink.com/content/l23864w241673530/}{Zeits. f. Physik 10 (1922) 377}. 

\bibitem{Lemaitre}G. Lema{\^i}tre, \emph{Un Univers homog{\`e}ne
de masse constante et de rayon croissant rendant compte de la vitesse
radiale des n{\'e}buleuses extra-galactiques}, Annales de la Societe
Scientifique de Bruxelles, A 47 (1927) 49 {[}\href{http://adsabs.harvard.edu/abs/1927ASSB...47...49L}{NASA ADS}{]}. 

\bibitem{Hubble}E. Hubble, \emph{A relation between distance and
radial velocity among extra-galactic nebulae}, \href{http://www.pnas.org/content/15/3/168}{PNAS 15 (1929) 168}.

\bibitem{Zel'dovich}Ya. B. Zel'dovich, \emph{The cosmological constant
and the theory of elementary particles}, \href{http://iopscience.iop.org/0038-5670/11/3/A13/}{Sov. Phys. Usp. 11 (1968) 381}.

\bibitem{Weinberg-1}S. Weinberg, \emph{The cosmological constant
problem}, \href{http://rmp.aps.org/abstract/RMP/v61/i1/p1_1}{Rev. Mod. Phys. 61 (1989) 1}.

\bibitem{acc:1998}A. G. Riess et al., \emph{Observational Evidence
from Supernovae for an Accelerating Universe and a Cosmological Constant},
\href{http://iopscience.iop.org/1538-3881/116/3/1009/}{Astron. J. 116 (1998) 1009}
{[}\href{http://arxiv.org/abs/astro-ph/9805201}{astro-ph/9805201}{]}. 

\bibitem{acc:1999}S. Perlmutter et al., \emph{Measurements of $\Omega$
and $\Lambda$ from 42 High-Redshift Supernovae}, \href{http://iopscience.iop.org/0004-637X/517/2/565/}{Astrophys. J. 517 (1999) 565}
{[}\href{http://arxiv.org/abs/astro-ph/9812133}{astro-ph/9812133}{]}.

\bibitem{Weinberg-2}S. Weinberg, \emph{The Cosmological Constant
Problems (Talk given at Dark Matter 2000, February, 2000)}, {[}\href{http://arxiv.org/abs/astro-ph/0005265}{astro-ph/0005265}{]}. 

\bibitem{Guth}A. H. Guth, \emph{Inflationary Universe:  A possible
solution to the horizon and flatness problems}, \href{http://prd.aps.org/abstract/PRD/v23/i2/p347_1}{Phys. Rev. D 23 (1981) 347}.

\bibitem{Linde}A. Linde, \emph{Inflationary Cosmology}, \href{http://www.springerlink.com/content/7131q2812376q36h/}{Lect. Notes Phys. 738 (2008) 1}
{[}\href{http://arxiv.org/abs/0705.0164}{arXiv:0705.0164}{]}.

\bibitem{reheating-1}L. Kofman, A. Linde and A. Starobinsky, \emph{Reheating
after Inflation, }\href{http://prl.aps.org/abstract/PRL/v73/i24/p3195_1}{Phys. Rev. Lett. 73 (1994)  3195}\emph{
}{[}\emph{\href{http://arxiv.org/abs/hep-th/9405187}{hep-th/9405187}}{]};
L. Kofman, A. Linde and A. Starobinsky, \emph{Towards the Theory of
Reheating After Inflation, }\href{http://prd.aps.org/abstract/PRD/v56/i6/p3258_1}{Phys. Rev. D 56 (1997)  3258}\emph{
}{[}\emph{\href{http://arxiv.org/abs/hep-ph/9704452}{hep-ph/9704452}}{]}. 

\bibitem{reheating-2} Y. Shtanov, J. Traschen and R. Brandenberger,
\emph{Universe reheating after inflation}, \href{http://prd.aps.org/abstract/PRD/v51/i10/p5438_1}{Phys. Rev. D 51 (1995) 5438}
{[}\href{http://arxiv.org/abs/hep-ph/9407247}{hep-ph/9407247}{]}.

\bibitem{reheating-3} B. A. Bassett, S. Tsujikawa and D. Wands, \emph{Inflation
Dynamics and Reheating}, \href{http://rmp.aps.org/abstract/RMP/v78/i2/p537_1}{Rev. Mod. Phys. 78 (2006)  537}
{[}\href{http://arxiv.org/abs/astro-ph/0507632}{astro-ph/0507632}{]}.

\bibitem{Moss}I. G. Moss, \emph{Primordial inflation with spontaneous
symmetry breaking}, \href{http://www.sciencedirect.com/science/article/pii/0370269385905702}{Phys. lett. 75 (1985) 3218}. 

\bibitem{Berera et al.}A. Berera and L.-Z. Fang, \emph{Thermally
induced density perturbations in the inflation era}, \href{http://prl.aps.org/abstract/PRL/v74/i11/p1912_1}{Phys. Rev. Lett. 74 (1995) 1912}
{[}\href{http://arxiv.org/abs/astro-ph/9501024}{astro-ph/9501024}{]};
A. Berera, I. G. Moss and R. O. Ramos, \emph{Warm Inflation and its
Microphysical Basis}, \href{http://iopscience.iop.org/0034-4885/72/2/026901/}{Rept. Prog. Phys. 72 (2009) 026901}
{[}\href{http://arxiv.org/abs/0808.1855}{arXiv:0808.1855}{]}.

\bibitem{dissipation 1}M. Bastero-Gil, A. Berera and R. O. Ramos,
\emph{Dissipation coefficients from scalar and fermion quantum field
interactions}, \href{http://iopscience.iop.org/1475-7516/2011/09/033/}{JCAP 1109 (2011) 033}
{[}\href{http://arxiv.org/abs/1008.1929}{arXiv:1008.1929}{]}.

\bibitem{dissipation 2}I. G. Moss and C. Xiong, \emph{Dissipation
coefficients for supersymmetric inflationary models}, {[}\href{http://arxiv.org/abs/hep-ph/0603266}{hep-ph/0603266}{]}.

\bibitem{entropy 1}C. A. Egan and C. H. Lineweaver, \emph{A Larger
Estimate of the Entropy of the Universe}, \href{http://iopscience.iop.org/0004-637X/710/2/1825/}{Astrophys. J. 710 (2010) 1825}
{[}\href{http://arxiv.org/abs/0909.3983v3}{arXiv:0909.3983v3}{]}.

\bibitem{entropy 2}P. Frampton et al., \emph{What is the entropy
of the Universe?} \href{http://iopscience.iop.org/0264-9381/26/14/145005/}{Class. Quant. Grav. 26 (2009) 145005}
{[}\href{http://arxiv.org/abs/0801.1847v3}{arXiv:0801.1847v3}{]}.

\bibitem{BND_1997}M. Fukugita, C. J. Hogan and P. J. E. Peebles,
\emph{The Cosmic Baryon Budget}, \href{http://iopscience.iop.org/0004-637X/503/2/518/}{Astrophys. J. 503 (1998) 518}
{[}\href{http://arxiv.org/abs/astro-ph/9712020}{astro-ph/9712020}{]}.

\bibitem{WMAP7}E. Komatsu et al., \emph{Seven-Year Wilkinson Microwave
Anisotropy Probe (WMAP) Observations: Cosmological Interpretation},
\href{http://iopscience.iop.org/0067-0049/192/2/18/}{Astrophys. J. Suppl. 192 (2011) 18}
{[}\href{http://arxiv.org/abs/1001.4538}{arXiv:1001.4538}{]}.

\bibitem{acc:2011}A. G. Riess et al.\emph{, A $3\%$ Solution: Determination
of the Hubble Constant with the Hubble Space Telescope and Wide Field
Camera $3$, }\href{http://iopscience.iop.org/0004-637X/730/2/119/}{Astrophys. J. 730 (2011) 119}
{[}\href{http://arxiv.org/abs/1103.2976}{arXiv:1103.2976}{]}.

\bibitem{Thermal_inflation-1}E. Gunzig, R. Maartens and A. V. Nesteruk,
\emph{Inflationary cosmology and thermodynamics}, \href{http://iopscience.iop.org/0264-9381/15/4/014/}{Class. Quant. Grav. 15 (1998) 923}
{[}\href{http://arxiv.org/abs/astro-ph/9703137}{astro-ph/9703137}{]}.

\bibitem{Thermal_inflation-2}A. V. Nesteruk, \emph{Inflationary Cosmology
with Scalar Field and Radiation}, \href{http://www.springerlink.com/content/x28153474033g182/}{Gen. Rel. Grav. 31 (1999) 983}
{[}\href{http://arxiv.org/abs/gr-qc/9905105}{gr-qc/9905105}{]}. 

\bibitem{Prigogin}I. Prigogine, J. Geheniau, E. Gunzig and P. Nardone,
\emph{Thermodynamics and Cosmology}, \href{http://www.springerlink.com/content/x4t2824k232m4032/}{Gen. Rel. Grav. 21 (1989) 767}.

\bibitem{PFC}V. Mukhanov, Physical Foundations of Cosmology, \href{http://prp.contentdirections.com/mr/cupress.jsp/doi=10.2277/0521563984}{Cambridge University Press (Cambridge),2005}. 

\bibitem{early_universe}E. W. Kolb and M. S. \emph{Turner, The Early
Universe}, \href{http://www-spires.slac.stanford.edu/spires/find/hep/www?irn=2256878}{Front. Phys. 69:1-547, 1990}. 

\bibitem{Neu-1}S. Hannestad and J. Madsen, \emph{Neutrino decoupling
in the early Universe}, \href{http://prd.aps.org/abstract/PRD/v52/i4/p1764_1}{Phys. Rev. D 52 (1995) 1764}
{[}\href{http://arxiv.org/abs/astro-ph/9506015}{astro-ph/9506015}{]}.

\bibitem{YCCHEN}Y.-C. Chen, \emph{The} \emph{creation of radiation
and the relic of inflaton potential,} {[}\href{http://arxiv.org/abs/1109.6612}{arXiv:1109.6612}{]}.

\end{thebibliography}
\end{document}